\documentclass[preprint]{emulateapj}

\newcommand{\lya}{Ly$\alpha$}
\newcommand{\hb}{H$\beta$}
\newcommand{\ha}{H$\alpha$}
\newcommand{\flux}{erg s$^{-1}$ cm$^{-2}$}
\newcommand{\oii}{[\ion{O}{2}]}
\newcommand{\oiii}{[\ion{O}{3}]}
\newcommand{\iacs}{$I_{814}$} 
\newcommand{\clm}{Paper I} 
\newcommand{\dressler}{Paper II}

\begin{document} 
\title{The faint-end slope of the redshift 5.7 \lya\ luminosity function\altaffilmark{1,2}}
\author{Alaina L. Henry\altaffilmark{3}, Crystal L. Martin\altaffilmark{3}, Alan Dressler\altaffilmark{4},  Marcin Sawicki\altaffilmark{5} \& Patrick McCarthy\altaffilmark{4}}

\altaffiltext{1}{ Some of the data presented herein were obtained at the W.M. Keck Observatory, which is operated as a scientific partnership among the California Institute of Technology, the University of California and the National Aeronautics and Space Administration. The Observatory was made possible by the generous financial support of the W.M. Keck Foundation.}
\altaffiltext{2}{This paper includes data gathered with the 6.5 meter Magellan Telescopes located at Las Campanas Observatory, Chile.}
\altaffiltext{3}{Department of Physics, University of California, Santa Barbara, CA 93106; ahenry@physics.ucsb.edu}
\altaffiltext{4}{Carnegie Observatories, 813 Santa Barbara Street, Pasadena, CA 91101}
\altaffiltext{5}{Department of Astronomy and Physics, Saint Mary's University, Halifax, NS B3H 3C3, Canada}

\begin{abstract}  
Using new Keck DEIMOS spectroscopy, we examine the origin of the steep number counts of 
 ultra-faint emission-line galaxies
recently reported by Dressler et al.\ (2011).   We confirm
six \lya\ emitters (LAEs), three of which have significant asymmetric line profiles with prominent wings 
extending 300-400 km s$^{-1}$
redward of the peak emission.    
With these six LAEs, we revise our previous estimate of the number of 
faint LAEs in the Dressler et al.\ survey.  Combining these data with the density of bright LAEs in the 
Cosmic Origins Survey and  Subaru Deep Field  provides
the best constraints to date on the redshift 5.7 LAE luminosity function (LF). 
Schechter function parameters,
$\phi^* = 4.5 \times 10^{-4}$ Mpc$^{-3}$, $L^* = 9.1\times10^{42}$ erg s$^{-1}$, and  $\alpha= -1.70$, are estimated using a
maximum likelihood technique with a model for slit losses.  To place this result
in the context of the UV-selected galaxy population, we investigate how various parameterizations of
the \lya\ equivalent width distribution,  along with the measured UV-continuum LF,  affect shape and normalization of the \lya\ LF.
The nominal model, which uses $z\sim6$ equivalent widths from the literature,  falls 
short of the observed space density of LAEs at the bright end, possibly indicating a need for higher equivalent widths.  
 This parameterization of the equivalent width distribution implies that as many as
50\% of our faintest LAEs should have $M_{UV} > -18.0$, rendering them undetectable in even the deepest
{\it Hubble Space Telescope} surveys at this redshift. Hence, ultra-deep emission-line surveys find
some of the faintest galaxies ever observed at the end of the reionization epoch. Such faint galaxies
likely enrich the intergalactic medium with metals and maintain its ionized state.  Observations of these objects provide a 
glimpse of the building blocks of present-day galaxies at an early time.

\end{abstract}

\section{Introduction} 
Observations of galaxies in the post-reionization epoch ($z\sim4-6$) measure the growth of young 
objects in an early Universe.  Over these redshifts, galaxies assemble into the active systems that we observe during the peak of
cosmic star-formation at $z\sim2-3$.  These galaxies drive winds, which play a critical role in the 
metal enrichment of the intergalactic medium (IGM;  \citealt{Opp09, clm10}).  Additionally, observations of galaxies in the
 post-reionization era provide a  baseline for understanding the earliest star formation and reionization 
 of the IGM at $z>7$.

Studies of both Lyman Break Galaxies (LBGs; \citealt{Bouwens07}) and Ly$\alpha$ emitters (LAEs;
 \citealt{Rhoads00, Shimasaku, Gronwall, Ouchi08, Ouchi10, Hu10, Kashikawa11})  in the post-reionization era 
 provide complementary views  of early star formation and galaxy evolution.
   The escape of \lya\ photons from the interstellar media of galaxies is strongly influenced by galactic outflows.   Typically, observable emission results when photons are redshifted out
  of resonance with the systemic velocity of the host galaxy \citep{Shapley03, Steidel10, Steidel11}.  As a result, the comparison 
  between 
  LAEs and LBGs contains information about star formation driven galactic winds.   For example,  the space 
density of UV-luminous galaxies declines with increasing redshift at $z\sim4-6$, while \lya-- luminous galaxies maintain an 
approximately constant
space density  \citep{Bouwens07, Ouchi08}. The direct implication of these results is that \lya\ photons more readily escape from 
galaxies at higher redshifts, possibly due to decreasing dust or the increased presence of gaseous outflows.   Nevertheless, current 
conclusions must be drawn from limited data, as narrowband imaging observations leave the faint-end of the \lya\ luminosity
 function (LF) unconstrained.

Looking to the end of reionization, it is still debated whether galaxies
produced enough ionizing photons to prevent recombination in the IGM. It is often
 argued that galaxies fainter than those present in the deepest {\it Hubble Space Telescope} (HST) 
 surveys are needed to produce enough ionizing photons to balance recombinations in the 
 IGM \citep{YH04,Salvaterra, Finlator11}.   
Because of this requirement for fainter objects, LAE surveys at this redshift are useful, 
as they can identify galaxies that cannot be detected by other means.   
 As an example, an LAE with $L=10^{42}$  erg s$^{-1}$ and rest-frame equivalent 
 width $W_0>$  80 \AA\ would imply a UV-luminosity of  $M_{UV} > -18.0 $ --- a  
 level too faint to be detected in the  Hubble Ultra Deep Field \citep{Bouwens06}.    Identifying very faint LAEs 
 can shed new light on the ionizing photon budget by including sources with continuum luminosities that have never been 
 probed in the epoch of reionization.

 In addition,
 studies of LAEs in the post-reionization 
 era are crucial if we aim to determine the neutral hydrogen fraction from \lya\ emission.   
 Measurements of \lya\ equivalent widths, line profiles, clustering, and the LF will all be 
 impacted  by neutral hydrogen \citep{ME2000, Furlanetto2006, Dijkstra07_lineprof, Dijkstra07, McQuinn,Dayal, Schenker},
  but quantifying these effects 
 will require robust measurements in the post-reionization era  for comparison.    Because galaxies in a partially 
 neutral IGM  are thought to reside in 
 bubbles ionized by their own star-formation, luminosity dependent trends are expected \citep{Ono}.   By measuring only
 bright LAEs,  narrowband imaging provides an incomplete view of these star-forming galaxies.

Clearly, samples of faint LAEs  are required to better understand  galaxies
and the IGM at $z>4$.   To detect these objects, {\it spectroscopic} searches are 
necessary  to increase the contrast with the sky 
background.   Blind spectroscopic searches for LAEs have now successfully identified
 galaxies at   $3 < z < 6$ 
(\citealt{Santos, clm08}, hereafter \clm, \citealt{Rauch08, Sawicki08, Cassata}). 
 In Dressler et al. (2011, hereafter \dressler) we presented first results from a new 
 Multislit Narrowband Spectroscopic (MNS) survey for LAEs at
$z=5.7$.  In this search, we   reached unprecedented depths and detected 215 
LAE candidates.  The faintest objects, if they are confirmed, will 
have luminosities that have only been seen (at this redshift) in a few lensed galaxies \citep{Santos}.     
Only 31 (14\%) the LAE 
candidates in \dressler\ have fluxes bright enough to be selected through 
narrowband imaging ($1.0 \times 10^{-17}$ erg s$^{-1}$ cm$^{-2}$)\footnote{Ground-based 
narrowband imaging searches are limited by residual flat-fielding errors \citep{ST06}, so increased 
exposure times will not reveal fainter galaxies.}.

In \dressler,  we presented evidence that a high fraction of the LAE candidates are 
indeed LAEs, but spectroscopic confirmation 
 is still required.   In the present paper, we analyze  the first results from our 
 spectroscopic followup campaign.   With a new sample of 
 confirmed LAEs in hand, we provide the first  measurement of the faint-end 
 slope of the \lya\ LF at redshift 5.7.  We demonstrate that slit-losses can be
 robustly accounted for in blind-spectroscopic data by using  Monte Carlo 
 simulations within the framework of the maximum likelihood method.  In fact,  
 as we show in \S \ref{lyalf_mle}, this method is directly applicable to narrowband imaging data,  
 where uncertainties often result from a non-uniform narrowband throughput.

This paper is organized as follows:  in \S \ref{survey} we give a brief overview of 
the survey design.   In \S 3 we present our spectroscopic followup observations, 
and in \S 4 we discuss the
newly identified LAEs.   These LAEs are used, in \S 5, to infer the number of LAEs 
that should remain in our sample, and in \S 6 we 
derive the \lya\ LF by combining our data with those from the Subaru Deep Field 
(SDF)  and COSMOS \citep{Scoville}.   Prospects for improving  
the parameter constraints are discussed in \S 7.   The implications of this work, including 
comparisons to other measurements, and a discussion of LAE properties can be found 
in \S 8.  Survey completeness is described in Appendix A.  Finally, in Appendix B we examine 
the foreground ($z\sim 0.25 - 1.2$) emission line galaxies and update our model for the interloper
 fraction that we previously presented in \dressler. 

Throughout this paper we use AB magnitudes, and a $\Lambda$CDM cosmology: 
 $H_0 = 70 ~{\rm   km s^{-1}~ Mpc^{-1}}$, $\Omega_{\Lambda} = 0.7$, $\Omega_M = 0.3$.  
 All volumes are given in comoving units, unless otherwise stated.

\section{Survey Design \& Followup Strategy} 
\label{survey}
The LAE candidates that we aim to confirm in this paper are selected from our spectroscopic
 survey, which is presented fully in  \dressler.    In summary, the search uses the Inamori-Magellan
  Areal Camera and Spectrograph (IMACS; \citealt{Dressler06}) on the Magellan-Baade Telescope. 
   To obtain blind, narrowband spectra we use a ``venetian blind'' slit-mask (see Figure 2 of \clm\ for a schematic) and a  
   134 \AA-wide (FWHM) blocking filter centered in the 8200\AA\ OH-free atmospheric window. 
The slits on this mask cover an area of 55 arcmin$^2$, corresponding to a volume of 
 1.5$\times 10^{4}$ Mpc$^{3}$ for \lya\ at redshift 5.7.   Emission lines that are detected are as 
 faint as   2.5$\times 10^{-18}$ \flux.   Two fields were surveyed: one in the center of  the COSMOS
  field \citep{Scoville},   and the other in the  Las Campanas Infrared Survey (LCIRS) 15H
   field \citep{marzke}.  In this paper we focus on the COSMOS field, as the 15H field has only
    limited 
spectroscopic followup at this time.  A more complete luminosity function, with data 
from both fields combined, will be the subject of a future paper. 

\begin{figure*} 
\begin{center} 
\includegraphics[width=7.0in]{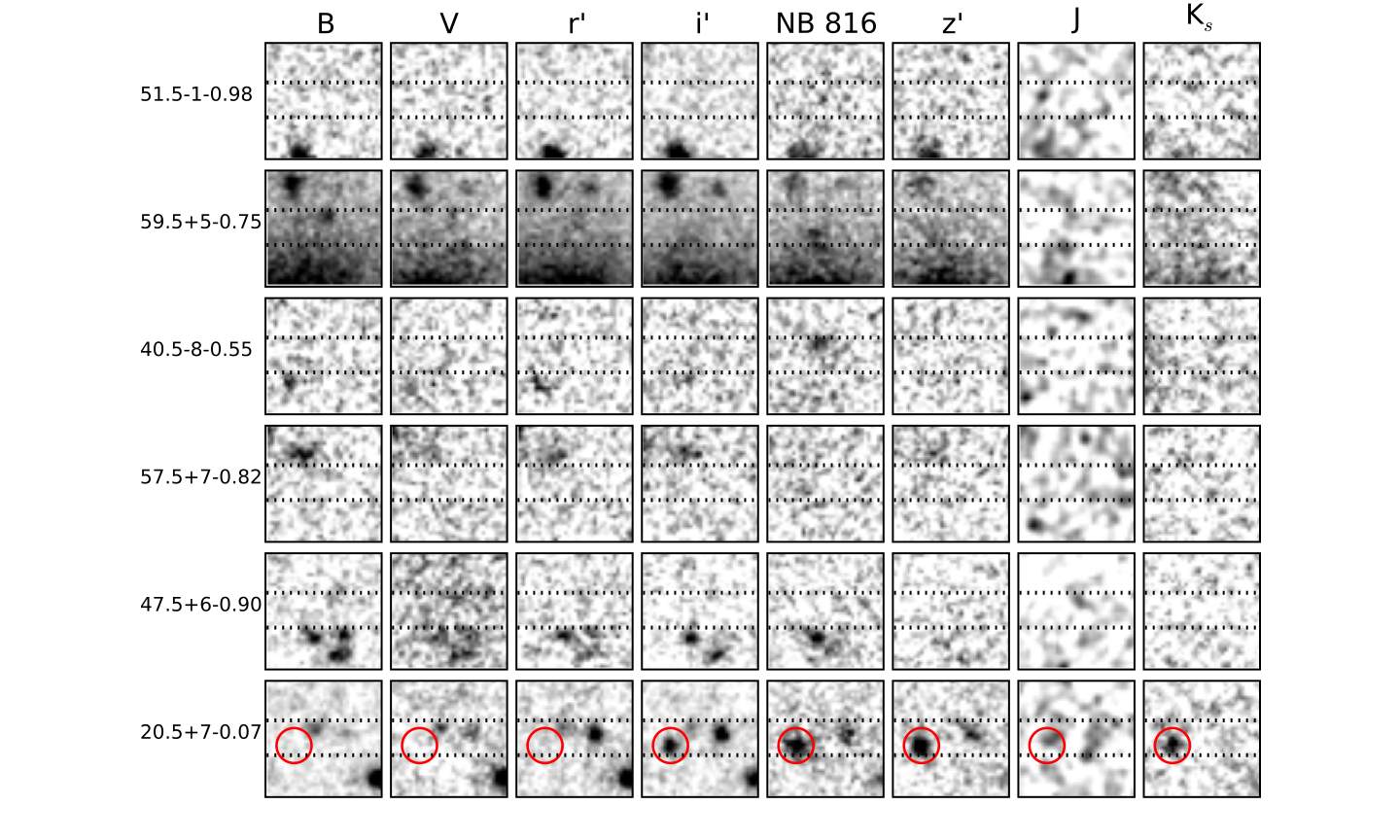}
\end{center}
\caption{Thumbnail images centered on the locations of the confirmed LAEs presented in
 \S \ref{confirmed_sec}.  Image are 5\arcsec\
on a side, and the dotted lines show the location of our slit.  The  Subaru (B, V, r, i, NB  816, and z), 
UKIRT (J), and CFHT ($K_s$) images are taken from the COSMOS image cutout service,  
hosted by the NASA/IPAC Infrared Science Archive.
  The $i-$ dropout near 20.5+7-0.07 (see text) is circled in red.  } 
\label{stamps_array}
\end{figure*} 

We identified 105 LAE candidates\footnote{We have added one object that was dropped 
from the sample in \dressler, bringing the number of LAE candidates
from 104 to 105.}  in the IMACS search  of the COSMOS field. These objects are seen 
as single emission lines with no 
detected continuum (or, in a few cases, a possible weak continuum detection only to the red of the line). 
 An additional 130 objects with blue continuum or multiple emission lines\footnote{Multiple emission lines
  can be present in the 134 \AA\ of search spectrum when we detect \ha\ and 
  [\ion{N}{2}] $\lambda$6583 \AA, or \oiii\  $\lambda \lambda$4959, 5007 \AA.}  
   were identified as foreground galaxies in the search data.  

Identifying the LAE candidates as genuine LAEs or foreground galaxies is not trivial.  
Since  the positions of galaxies within the
slits  are not known, a large area on the sky must be searched for an optical 
counterpart.  As a result, the normal approach of requiring an optical non-detection 
to distinguish foreground galaxies from LAEs
is not applicable.   
This ambiguity is illustrated in Figure \ref{stamps_array}, which shows broadband and 
narrowband thumbnail images of our six confirmed LAEs (which we will discuss in \S \ref{confirmed_sec}). 
  In most cases there are no objects falling within the slits at the spatial position of the LAEs, but in 
some cases there are objects nearby or outside the slits.  Similarly,  only 37 of the 105 LAE candidates have 
entries 
within a 2\arcsec\ radius in the COSMOS photometric redshift catalog.\footnote{The photometric redshifts 
used in this paper  are taken from the COSMOS Photometric Redshift Catalog (Fall 2008; version 1.5), 
which can be found at  http://irsa.ipac.caltech.edu/Missions/cosmos.html.}   Of these, 20 have photometric 
redshifts that place an emission line in our 8200\AA\ bandpass and 17 do not.

 Clearly, the faintness of our 
LAE candidates means that spectroscopic followup is required to determine if the detected emission line is \lya, or 
an emission line from a foreground galaxy. 
Under this approach, we can detect other emission lines that identify the galaxies as 
foreground emitters, and also observe the ``smoking gun'' 
signature of \lya\ emission:  an asymmetric line profile.    In the sections below we describe
 our spectroscopic followup observations.  
The slits used to observe LAE candidates always have the same width,  coordinates, and 
position angle (PA) as the search slits, regardless of potential optical 
counterparts.

\section{Observations}

\begin{deluxetable*} {ccccccccc}
\tabletypesize{\scriptsize}
\tablecolumns{9}
\tablecaption{Observation Summary} 
\tablehead{
\colhead{Date} & \colhead{Instrument}  &  \colhead{Mask Name}  & \colhead{Mask RA}  & \colhead{Mask Dec} &\colhead{ Mask PA} & \colhead{Exposure Time}  & \colhead{Seeing}  & \colhead{Conditions }  \\
&  &    & & & & \colhead{(hours)}  & \colhead{(\arcsec)}  &
} 
\startdata
23-26 March 2009 & IMACS & 10hcon09 & 10:00:42.90  &  02:11:00.0 & 90 & 16.0  & 0.35-0.8 & Clear \\
7 March 2010	& DEIMOS & A  &   10:00:29.19  &  02:09:33.7  & 95  & 6.3	& 0.9-1.6  & Clear  \\
8 March 2010   &  DEIMOS & B &     10:00:22.10 &   02:01:38.0 &   95	& 6.0 	& 1.2-1.7 	& Cirrus \\
27 January 2011 & DEIMOS &  D &  10:00:22.97 &    02:09:28.9  & 85 	& 6.5		&	0.6	& Clear  \\
28 January 2011 &  DEIMOS &F &    10:01:11.46 &     02:10:27.4 & 106   	& 6.3		& 0.8		& Clear  
\enddata
\label{obstable} 
\tablecomments{Coordinates and position angle (PA) are given for each observed mask.    The slit PA, which differs
from the mask PA given here, was 90 degrees for every object.  Slit-widths were 1.5\arcsec.}
\end{deluxetable*} 

\label{specfollow}
\subsection{IMACS Spectroscopic Followup} 
Spectroscopic followup was carried out on March 23-26 2009, using the Inamori  
Magellan Areal Camera (IMACS; \citealt{Dressler06}) on the 
Magellan Baade Telescope.  Conditions were clear, and the average seeing 
over the three nights was 
0.6\arcsec.  A slit width of 1.5\arcsec\ was used to match the 2008 search observations, 
and the 200 line mm$^{-1}$ grating 
provided a  2 \AA\ pixel$^{-1}$ scale to go along with the 0.2\arcsec\ pixel$^{-1}$  scale of the IMACS f/2 camera.  
The full spectral range covered approximately 4000 - 9000 \AA. 

 The data were reduced using the COSMOS  (Carnegie Observatories System for Multi Object Spectroscopy) 
 software package\footnote{http://obs.carnegiescience.edu/Code/cosmos}.    These spectra allowed us to  
 identify  \ha,  \oiii, and \hb\ emission from  galaxies.  
However, at this S/N and resolution
neither the \oii\ doublet nor the \lya\ line profile can be resolved.   Therefore, both 
\oii\ and \lya\ emitters appear as single 
emission lines in our IMACS followup spectroscopy.   These single line objects are 
prioritized for higher-resolution spectroscopic
followup with DEIMOS (see below).

\subsection{DEIMOS Spectroscopic Followup}
Medium resolution followup spectroscopy with DEIMOS \citep{faber_deimos}   on Keck II was carried
 out on March 7-8, 2010 and January 27-28, 2011.  Further observations scheduled March 3-4 2011 were
  lost entirely to poor weather.   In addition to the high-priority single line emitters from our IMACS followup, 
   the slit-masks included LAE candidates which were not targeted for IMACS followup, and those that were 
   targeted but not recovered.      (Ultimately, it was not unusual for objects that we failed to recover in IMACS
    followup to be robustly detected under good conditions with DEIMOS.) 
 The observing conditions and mask positions are listed in Table \ref{obstable}.     In the 2010 observations, 
 poor seeing impeded our ability to recover the faintest sources and to distinguish \oii\ from \lya\ in low 
 S/N emission line detections.    Consequently, the optimal mask design in 2011 included substantial overlap
  with the previous year.   Two  LAEs identified on the March 2010 mask A  (51.5-1-0.98 and 59.5+5-0.75) were 
  also included on mask D in 2011.

Observations were made with the 830G grating, using slit-widths, position angles, and 
locations matched to the search data.  The data were reduced using the
DEEP2 DEIMOS Data Pipeline\footnote{http://astro.berkeley.edu/$\sim$cooper/deep/spec2d}, with an
 updated optical model for the 830G grating (P. Capak, private communication).  The January 2011 
 data were flux calibrated using observations of spectrophotometric standard stars, taken through a 
 1.5\arcsec\ slit at the parallactic angle.   The stars
used were G191B2B, GD50, Feige 66, Feige 67, and Hz 44 \citep{Oke, MG90}, with reference data 
taken from the ESO spectrophotometric standard star 
database\footnote{http://www.eso.org/sci/observing/tools/standards/spectra.html}.  
Sensitivity functions derived from each of these stars agree to within 12\%, and when 
Feige 66 is excluded agreement is better than 3\%. 
 In addition to the default pipeline extraction, the January 2011 spectra are extracted in 1\farcs8 apertures.  
This aperture size was chosen to match the  spatial extraction in \dressler, and is appropriate given
 the good seeing in both the search data and January 2011 followup. 
No flux calibration  was applied to the March 2010 data because variable seeing, at times larger than 
2\arcsec, and cirrus during some of the observations, implied significant count-rate 
differences from frame to frame.

Line fluxes were measured from the extracted January 2011 spectra, and compared 
to the fluxes measured from the IMACS search data (and used
in \dressler).   The root mean square deviation is approximately 50\%, even for bright sources.  
This scatter likely results from the fact that the LAE candidates are not necessarily 
centered in the slits, so small astrometric errors or different seeing will induce varied slit-losses from one observation to 
the next.  
In \S \ref{lyalf} we take this scatter into account when we derive the luminosity function.  
For consistency, all line fluxes in this paper are taken from the search data.

\section{Confirmed LAEs} 
\label{confirmed_sec}

\begin{figure*} 
\plotone{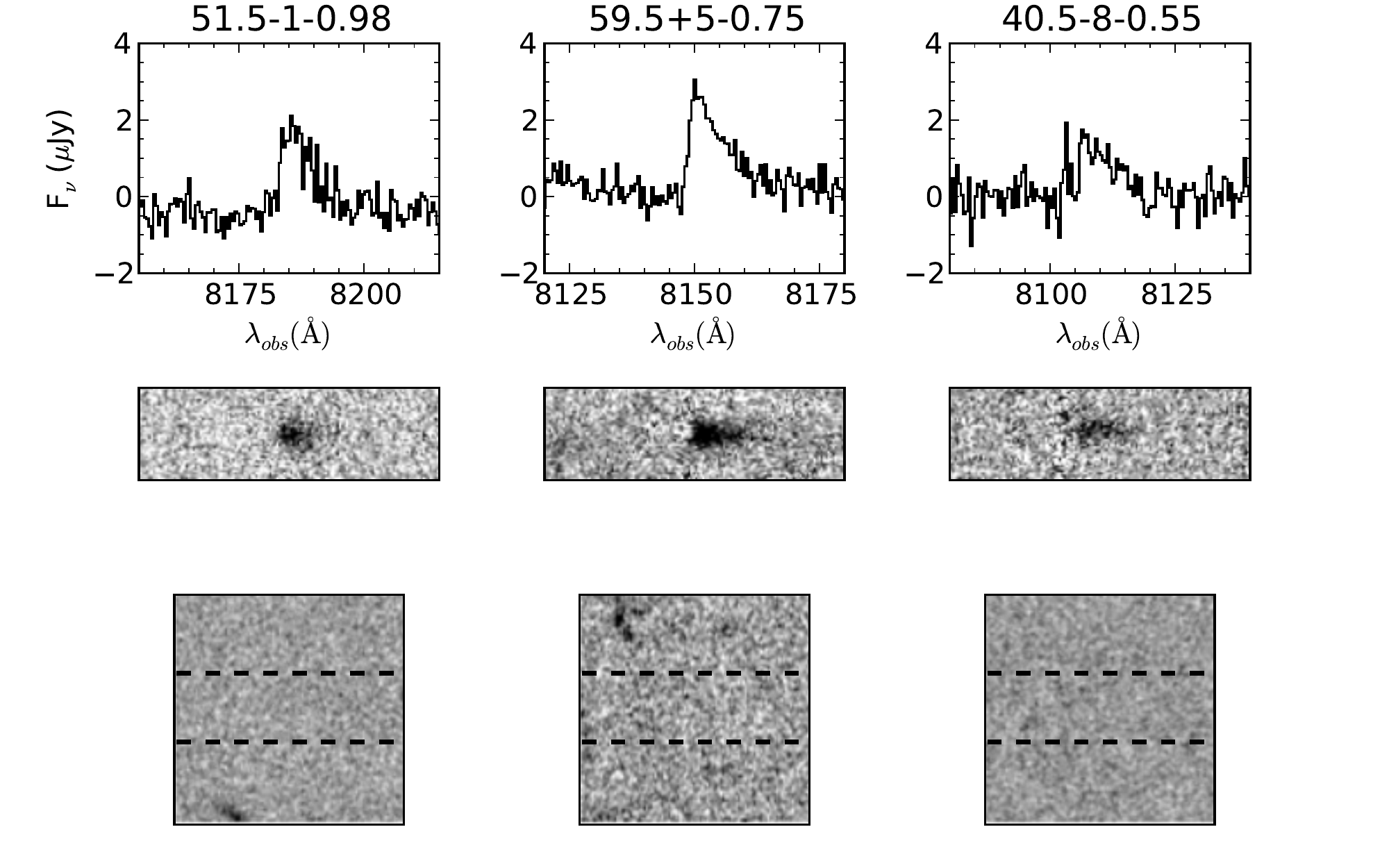} 
\caption{Spectra of high confidence, confirmed LAEs are shown in one dimension (top row) and
 two dimensions (middle row; dispersion in the horizontal direction).  The bottom row 
 shows 5\arcsec\  \iacs\ cutouts from the COSMOS ACS observations \citep{Koekemoer}.    The detection threshold in these images depends on
 the sizes of galaxies, but we can estimate that undetected sources have \iacs\ $ \ga 27$ \citep{Taniguchi, Henry09}.  The dashed 
lines show the location of our 1\farcs5-wide slits. } 
\label{montage1} 
\end{figure*}

Six LAEs are identified from the DEIMOS observations. Their coordinates, sizes, luminosities, velocity widths,  
and redshifts  are listed in Table \ref{LAEdata}.
 In Figure \ref{montage1} we 
show three cases of high quality \lya\ spectra, where the asymmetric line profile 
makes the identification unambiguous.      Little discussion is 
needed for these three LAEs, although we do note that the two brightest LAEs 51.5-1-0.98 and 59.5+5-0.75 
are clearly spatially resolved on mask D, where they were observed under 0\farcs6 seeing.    

\begin{figure*} 
\plotone{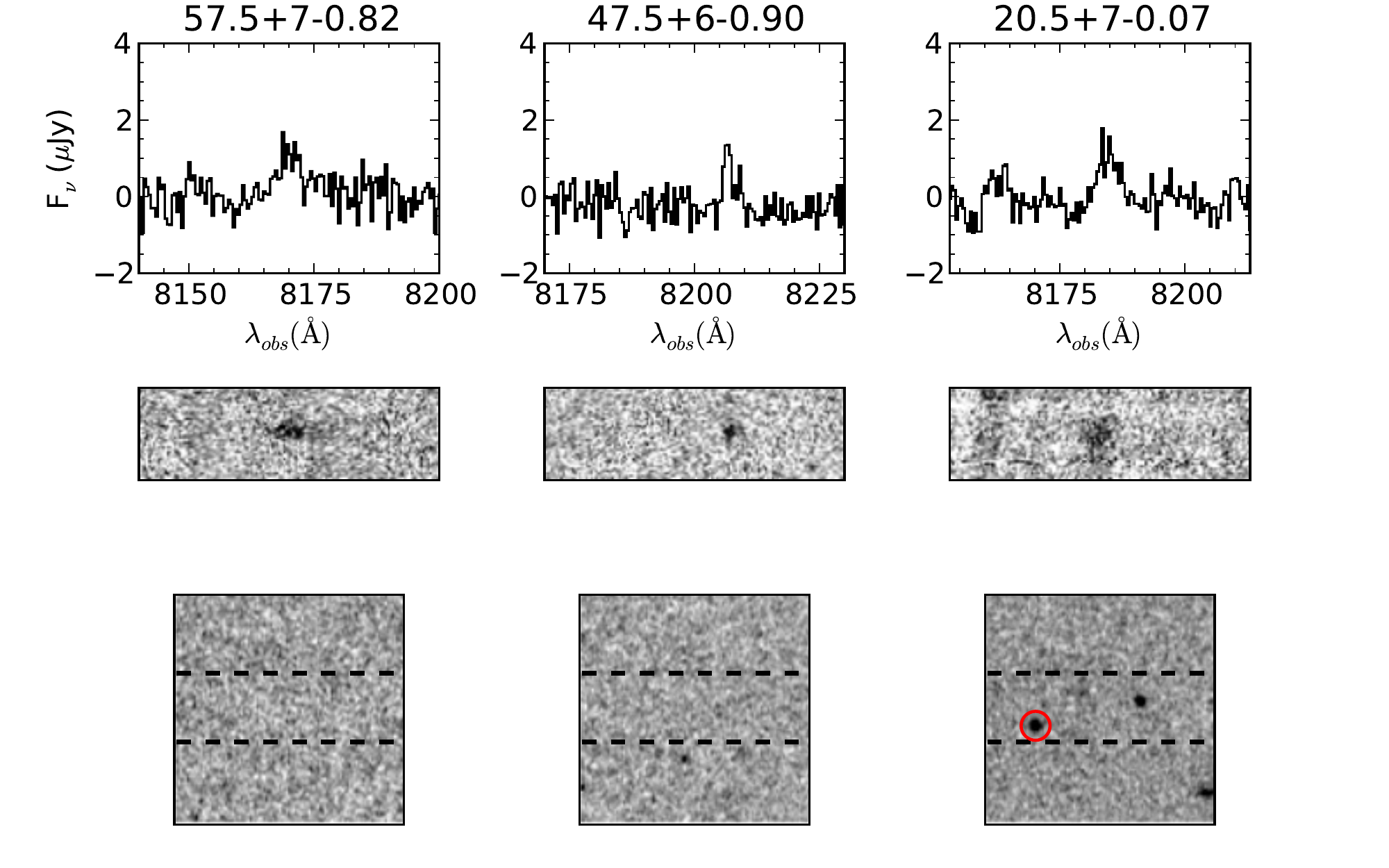}
\caption{ Same as Figure \ref{montage1}, but for less certain \lya\ identifications.  
The $i-$ dropout near 20.5+7-0.07 (see text) is circled in red.    } 
\label{montage2} 
\end{figure*}

When an emission line is not clearly asymmetric, more care is needed to determine if 
it is \lya.  Three such cases are identified, and are shown in 
Figure \ref{montage2} and listed in Table \ref{LAEdata}.       For these galaxies, other checks are 
possible.  First, we measure the S/N at the locations of \oiii $\lambda \lambda 4959, 5007$ \AA,  \oii $\lambda 3727$ \AA, 
and \hb\ for
the expected foreground populations.  In these cases, we find
 that \hb\ and \oiii\ $\lambda$4959 \AA\ are easily ruled out because other  stronger emission lines would be 
detected.  Likewise, the \oii\ doublet is  ruled out by the line profiles shown in  
Figure \ref{montage2}.   However, we can not rule out the case where we have 
detected \ha\ or \oiii\ $\lambda$5007\AA,  and all
other lines are below the detection threshold in our DEIMOS data.  While DEIMOS is efficient
 at 8000\AA, it is about a factor of two 
less sensitive at blue wavelengths where we would detect \oiii\ from 
an \ha\ emitter or \oii\ from an \oiii\ emitter. 
    Consequently, our spectra do not verify that  the emission lines 
    shown in Figure \ref{montage2} are genuine LAEs. 

\begin{deluxetable*} {ccccccccc}
\tabletypesize{\scriptsize}
\tablecolumns{9}
\tablecaption{Robust and Tentative LAEs} 
\tablehead{
\colhead{ID } & \colhead{RA\tablenotemark{a}}  & \colhead{Dec}  & \colhead{Flux} &\colhead{ Luminosity} & \colhead{Mask}  & \colhead{FWHM\tablenotemark{b}}  & \colhead{Redshift} & \colhead{$\sigma$\tablenotemark{c}}  \\
\colhead{} & \colhead{(J2000)} & \colhead{(J2000)}  & \colhead{($10^{-18}$ \flux)} &  \colhead{($10^{42}$ erg s$^{-1}$ ) }     &     &    \colhead{(\arcsec)}   &   & (km s$^{-1}$) }
\startdata
51.5--1--0.98  &10:00:49.766  &  02:10:06.448    &    $9.97 \pm 1.34$    &  3.72  & A\&D	&        1.13      &   5.730 & $163 \pm 70$ \\
40.5--8--0.55  & 10:01:35.446 &  02:07:22.911    &    $5.05 \pm 0.63$  & 1.83  	& F 	         &  	0.94     &  5.666  & $200 \pm 35$\\
57.5+7--0.82  &  09:59:59.604 &  02:11:37.826  &    $4.07  \pm 0.58$   &  1.51	& D		&	 0.69    & 5.717   &$187 \pm 132$ \\
59.5+5--0.75 & 10:00:11.686 &  02:12:07.171   &    $10.0 \pm 0.59$   &   3.68 	& A\&D      &      	  0.95    & 5.701  & $184 \pm 66$\\
47.5+6--0.90 &  10:00:07.145 &  02:09:07.808  &    $5.41  \pm 0.70$  &  2.03      &	D	& 	0.59	  &   5.749  & $<140$\tablenotemark{d} \\
20.5+7--0.07  & 10:00:00.691 &  02:02:24.636   &   $7.16  \pm 0.88$  &  2.67   & B		& 	1.22	&   5.730 &  $< 140$\tablenotemark{d} 
\enddata
\label{LAEdata} 
\tablenotetext{a}{The position accuracy in the R.A. direction (along the slit)  is 0.3\arcsec (RMS). }
\tablenotetext{b}{Spatial FWHM are seeing convolved and measured in the DEIMOS followup spectra.   (Accurate size measurements
are not possible with the low S/N in the search data.) The seeing on each mask is given in Table \ref{obstable}. } 
\tablenotetext{c}{Velocity width of a truncated Gaussian model fit to the line profile, as described in \S \ref{lineprof_sec}.} 
\tablenotetext{d}{These line profiles are consistent with being unresolved.} 
\end{deluxetable*}

The lack of any imaged counterparts {\it within} the 
slits 
for   57.5+70.82 and 47.5+6-0.90 (Figures \ref{stamps_array} and \ref{montage2}) suggests that emission lines may indeed by \lya. 
We next ask whether the implied  high equivalent width would rule out \ha\ or \oiii\ emission.  For an optical counterpart to be 
undetected in 
the Subaru $i$-band and the ACS \iacs\  it must be fainter than about 27th magnitude (AB; \citealt{Henry09, Taniguchi}). 
 This gives observed-frame equivalent width limits of  $W >  160$ and 210 \AA\ for these two candidates.   If, on the other hand, 
 the correct optical counterpart is one of the objects detected, but outside the slit, we would still expect a large equivalent width 
 because the true line 
 flux would be significantly attenuated by our slit.
 Nevertheless,  using the COSMOS NB816 catalogs which we derived in \dressler,  we find that 18-23\% of \ha\ and \oiii\ lines have 
 equivalent widths higher than these thresholds.  Since the extrapolated counts of \ha\ and \oiii\ emitters from \dressler\  are substantial, the high-equivalent width subset of these foreground populations
  could exist in similar numbers to the LAEs.   Nevertheless, 
 it is unlikely that \textit{both} 57.5+7-0.82 and  47.5+6-0.90 are foreground interlopers.

One LAE candidate in Figure \ref{montage2}  has not been discussed so far: 20.5+7-0.02. 
 Like the two other uncertain candidates, we can
not rule out \ha\ or \oiii\ from the spectrum alone.  (Although, we note that this object was observed
 under very poor seeing in March 2010, so future observations should confirm or rule out this LAE candidate.)   
 However, the \iacs\ postage stamp contains multiple objects.   We explore this further by examining
the multi-wavelength photometry from the COSMOS survey.  Thumbnail images are shown in Figure \ref{stamps_array};
 remarkably, an object
1\farcs3 east of our detected emission shows the appearance of an $i-$dropout 
(circled in Figures \ref{stamps_array} and \ref{montage2})\footnote{Although the
 $i-$dropout candidate in question was clearly inside our DEIMOS slit, no line 
 emission is detected.  This is not surprising, given the large UV-luminosity of this 
 galaxy ($M_{UV} \sim -22.0$ if it is at redshift  5.7), and the observation that \lya\ emission is not typical of
  such bright $i-$dropouts \citep{Stark11}.  Indeed, the NB816 photometry is consistent with no line emission.}.   
  If the 
 detected emission is ultimately confirmed to be \lya, we may have 
 observed a fainter companion galaxy.   However, the bluer foreground
galaxy lying near the center of the thumbnail image is more closely
 aligned with the detected emission, but its redshift is uncertain.  Due to its faintness, 
 this galaxy is not included in the COSMOS photometric redshift catalog.

In summary, we find three certain LAEs with clear asymmetric profiles and
 three plausible LAEs, where the blue sensitivity of our followup spectroscopy 
is insufficient to rule out all cases of \ha\ and \oiii\ $\lambda$5007\AA.   In the sections
 that follow we derive  \lya\ LFs for two cases: the ``3 LAE LF,''  where only
the robustly identified LAEs are included, and also the ``6 LAE LF'', where both the
 robust and plausible LAEs are included.

\section{LAE and Foreground Galaxy Identification Rates}   
\label{ratesec} 
In order to derive the \lya\ LF, we must address the question of how many LAEs remain 
in our survey that have not yet been spectroscopically confirmed.   To do this,  we supplement our IMACS 
and DEIMOS spectroscopy
with COSMOS photometric redshifts in order to identify as many foreground objects as possible.

To use the photometric redshifts, we must first  define what we refer to as a {\it foreground photometric match}.    As can be seen in Figure \ref{stamps_array},  nearby objects are present 
even around the robustly identified LAEs, so we can not identify foreground interlopers by the presence of optical counterparts. However, the situation is significantly improved  
with the COSMOS photometric redshifts \citep{Ilbert},  because the  galaxy
 templates contain 
emission lines, {\it and} the NB816 band is included in the photometric redshift derivation.     
 The consequence of this approach, is that-- if line emission is significant in the NB816 band-- the 
photometric redshifts become ``tuned''  to the precise redshifts of our foreground populations.  
For the NB816 filter, these
 redshifts are: $z = 1.17 - 1.2$ for \oii\ $\lambda$3727 \AA, $z=0.62-0.69$
  for \oiii $\lambda \lambda$4959, 5007 \AA, and \hb, and $z=0.20 - 0.25$ for \ha\ and
   [\ion{S}{2}] $\lambda \lambda$6716, 6731.  To find 
 photometric matches with all of our emission line galaxies, we
  search to a radius of 2\arcsec\ from the slit-center.  This gives 20/105 
  matches among the LAE candidates, and 77/130 matches among objects
   identified as foreground galaxies in the search data.   
With the caveat that this technique cannot 
distinguish \oiii\ $\lambda$5007 \AA\ from \oiii\ $\lambda$4959 \AA\ or  \hb , and
 possibly \ha\ from  [\ion{S}{2}] $\lambda \lambda$6716, 6731 \AA, spectroscopic followup shows that 
 photometric matches provide the correct redshift identification in 24/24 observed
   cases.\footnote{Many more foreground  objects were followed up 
   at the position of the photometric match and not 
the position of the IMACS search slit.  Here, we only count the objects 
where the followup slit was at the same position as the search slit.  }

In Table \ref{counts1}  we list the counts of foreground galaxies, LAEs, and unidentified
 objects by flux bins.   Line fluxes are defined in the first column, and are 
established to correspond to 0.2 dex bins centered at  log$(L/{\rm erg~s^{-1}})  = 42.1, 42.3,  42.5...$ for \lya\ at 
redshift 5.7.     The total number of emission line galaxies in the survey is given in column two, and broken into 
LAE candidates and foreground galaxies (both from the search data) in columns three and four.   Column
 five lists all foreground emission lines that we have identified to date, including objects that were initially
  LAE candidates but have foreground 
photometric matches or spectroscopic identifications. 
    Columns six and seven  list the counts of confirmed LAEs for the six and three LAE cases, respectively.   
    The final column (eight) lists the remaining  unidentified LAE candidates.    The sum of these 
    unidentified object counts plus the six likely confirmed LAEs represent an absolute upper limit 
    on the counts of faint  LAEs in COSMOS.
  For accounting purposes,  it should be noted that this conservative upper limit,  plus the identified
  foreground galaxies column are  equal to the $N_{obj}$ given in the second column.  Additionally, 
  in these binned counts we have excluded ten emitters fainter than
   log($F$ / \flux) $= -17.57$ and two brighter than $-15.57$.

\begin{deluxetable*} {cccccccc}
\tabletypesize{\scriptsize}
\tablecolumns{8}
\tablecaption{LAE and Foreground Counts } 
\tablehead{
\colhead{log $F$} & \colhead{$N_{obj}$}  & \colhead{LAE candidates}  &\colhead{Search Data Foregrounds}
 & \colhead{Identified Foregrounds}  & \colhead{LAEs (all)}  &  \colhead{LAEs (secure)} & \colhead{Unidentified} 
}
\startdata
-17.47  &  42  &  32 &   10    & 19	 &      1 &0      &  22   \\
-17.27  &  54  &  26  &  28    & 39  	 &  	 2  & 1    &  13   \\
-17.07  &  45  &  26  &  19    & 33     &	 3   & 2   &   9  \\
-16.87  &  30  &  9   &   21   & 24     &       0   & 0  &  6   \\
-16.67  &  16  &  3  &  13    &	 14 	 & 	 0    &0  &   2   \\
-16.47  &  14  &  0    &  14    & 14     & 	 0    & 0 &   0  \\
-16.27  &  13  &  1     &  12    &  13    &	0    &  0 &  0 \\
 -16.07  &    6  &   0   &     6   &   6      &      0   & 0 &   0  \\
-15.87  &    2  &   0    &     2    &   2     &      0    & 0 & 0  \\
-15.67  &    1  &   0    &     1    &  1      &     0    &  0&   0 
\enddata
\label{counts1} 
\tablecomments{ Number counts are given for the various categories of objects in the search data, as well as
after followup observations.    Columns two through four are based strictly on the search data, while
the remaining columns take our spectroscopic followup results into consideration.  
The final column gives the number counts of LAE candidates which have no spectroscopic or photometric identification. } 
\end{deluxetable*} 

We next aim to infer, from our spectroscopy, the number of LAEs and foreground galaxies that we expect to 
find among the unidentified line emitters
in Table \ref{counts1}.   For this inference, we consider only the spectroscopic identifications of objects without photometric 
matches, as the unidentified objects, by definition, exclude these matches.    Additionally, we examine only the three 
faintest flux bins given in 
Table \ref{counts1}.   The brighter bins contain only small numbers of LAE candidates, and none of these objects that lack 
photometric matches have been observed with DEIMOS.   Table \ref{rates} quantifies these foreground and LAE 
identifications.     While the numbers of galaxies with spectra remain small, no specific population of foreground emitter  
or LAEs appears to dominate.   (Although we will discuss the significant presence of \hb\ emitters in Appendix \ref{fgsec}).   

In order to calculate the number of LAEs that are likely present among the unidentified objects, we must recognize 
that these objects consist of two sub-classes.  
First, there are 27 galaxies where the low-resolution IMACS followup confirmed 
a single emission line, but was unable to resolve the \oii\  doublet.
Followup spectroscopy with DEIMOS showed that these objects are predominantly
 \oii\ and \lya, but, to date, not all of them have been targeted for higher-resolution spectroscopy.  
 On the other hand, the many galaxies which
 have no IMACS followup are observed to be a mix of \lya\ and all of the other 
 bright rest-frame optical emission lines.   These two sub-categories of emission 
 line galaxies populate the number counts of unidentified emitters
in Table  \ref{counts1}, but the former has a higher LAE fraction.   
  Therefore, we split these counts, and give LAE fractions appropriate 
for each sub-category in Table \ref{rates}. 
  The total numbers of inferred LAEs  are calculated for both the 3 LAE LF and also the 6 LAE LF. 
 
\begin{deluxetable*} {cccccccccccc}
\tabletypesize{\scriptsize}
\tablecolumns{10}
\tablecaption{LAE and Foreground Rates  among line emitters with no photometric matches} 
\tablehead{
\colhead{log $F$} & \colhead{$N_{spec}$} & \colhead{No ID}  & \colhead{\lya\ (all) } & \colhead{\lya\ (secure)}  &  \colhead{\ha} &\colhead{\hb} & \colhead{[\ion{O}{3}]}  & \colhead{\oii}  & \colhead{Other\tablenotemark{a}} &\colhead{Inferred LAEs (all)} & \colhead{Inferred LAEs (secure) }  
}
\startdata
\cutinhead{Line Emitters with no IMACS followup} 
-17.47  &  8  & 19  & 1 &   0  &  1  & 2  &  1  & 2	 &      1      &  2.1 &   0  \\
-17.27  &  7  & 9  & 2  &  1   &  0 &  3    &  0     & 1  	 &  	 1 &  2.3  &   1.1    \\
-17.07  &  6  &  6 & 3  &  2 &  1 &  0    &  0    & 3     &	 1    &  1.8  &  1.2  \\ 
\cutinhead{IMACS confirmed single line emitters} 
-17.47  & 2   & 3  & 0 &   0  &  0  & 0  &  0  & 2	 &      0      &   0 &   0  \\
-17.27  & 2   & 4  & 1 &   1  &  0  & 1  &  0  & 0	 &      0      &   2 &   2  \\
-17.17  & 3   & 3  & 2 &   1  &  0  & 0  &  0  & 1	 &      0      &   2 &   1  \\
\cutinhead{Total LAEs (Confirmed and Inferred)} 
 -17.47  &    &   &  &    &    &   &    & 	 &         &   3.1 &  0    \\
-17.27  &    &   &  &     &    &   &    & 	 &         &   6.3 &  4.1    \\
-17.17  &     &    &  &     &   &   &   & 	 &          &   6.8 &   4.2  
\enddata
\label{rates}
\tablenotetext{a}{The other foreground galaxies which we identify are two H$\gamma$ emitters and a galaxy with blue continuum but an uncertain 
redshift.}  
\end{deluxetable*}

\section{The \lya\ luminosity function} 
\label{lyalf} 

\subsection{Direct Calculation}  
\begin{figure}
 \plotone{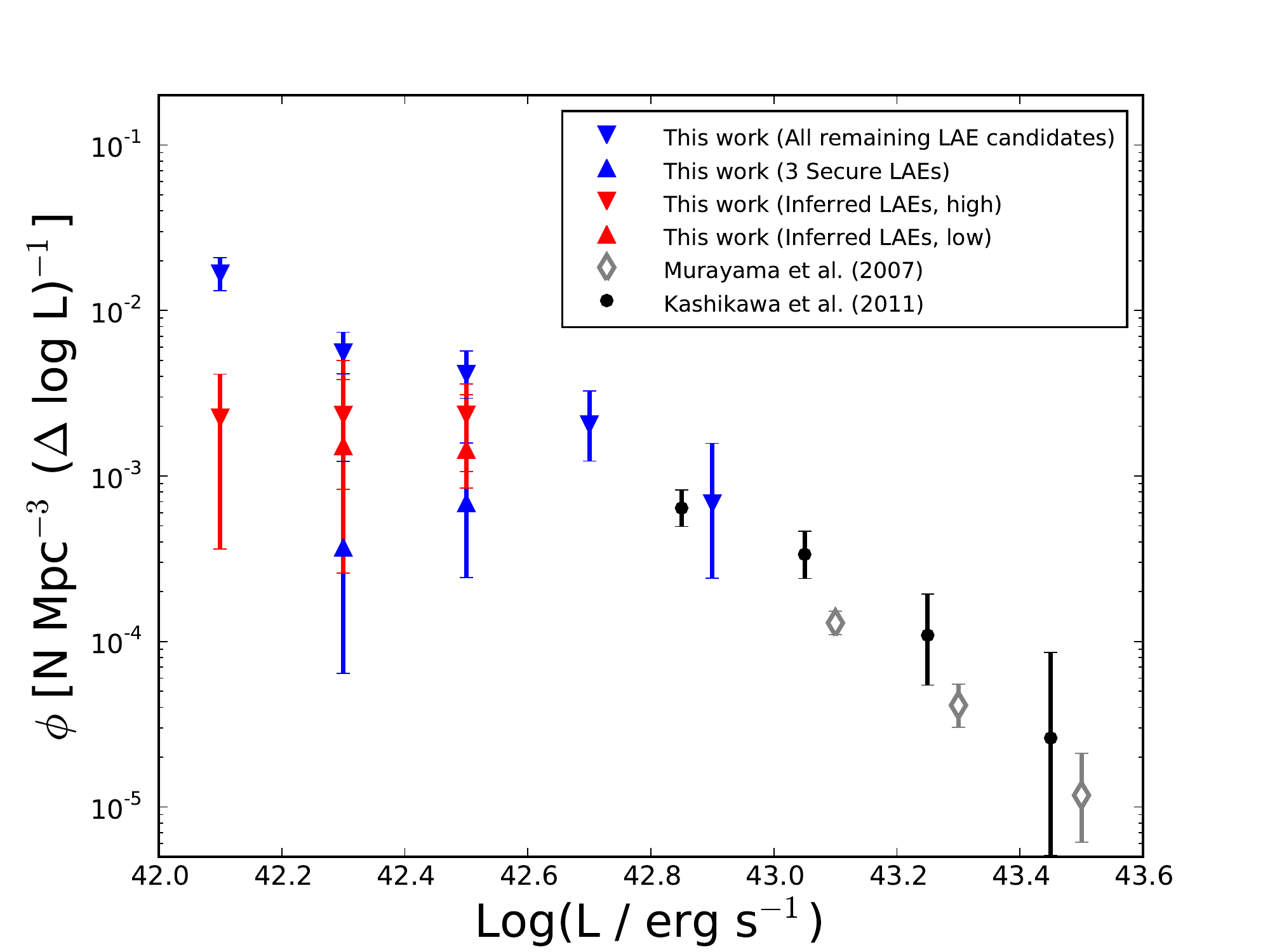}
 \caption{The discrete luminosity functions derived from 
 our blind spectroscopic sample  are shown in red (inferred from the  low, 
 3 LAE and high or 6 LAE samples).   The more conservative 
 limits are shown in blue.   For comparison, the bright-end data from the
  Subaru Deep Field (SDF;\citealt{Shimasaku, Kashikawa11}) and
   COSMOS \citep{Murayama} are shown.   The SDF data shown 
   here exclude the faintest candidates reported by 
 \cite{Kashikawa11}, where the spectroscopic confirmation 
 rate is low.}
 \label{lf_bins} 
  \end{figure}

With the counts of LAEs given in Tables \ref{counts1} and \ref{rates}, a 
direct derivation of the LF is straightforward.  The volume subtended 
by the slits in our search mask is 1.5$\times 10^{4}$ Mpc$^{3}$.  Correcting 
these counts for the detection completeness (Appendix \ref{comp_app}),
 we derive the conservative lower and upper limits (blue triangles).  Under 
 this approach, the lower limit is derived by including {\it only} the three robustly 
 identified objects.  The conservative upper limit, on the other hand, also 
 includes the three tentatively identified LAEs and all of the unidentified objects in 
 Table \ref{counts1}.   These discrete LFs are plotted in Figure \ref{lf_bins}. 
   No slit-loss corrections are applied at this stage;  they will be addressed \S  \ref{lyalf_mle}.  

In order to derive a meaningful constraint on the faint-end slope of the LF, 
data that cover the bright-end and the knee are needed.   
For this calculation, we draw on LAE catalogs available in COSMOS \citep{Murayama} and 
the Subaru Deep Field (SDF; \citealt{Kashikawa11},  K. Shimasaku,  private communication).    
While spectroscopic observations are more sensitive to low equivalent width emission,  
no systematic correction is needed, because the contribution from these sources is negligible. 
   No LAE candidates are seen with (convincingly) detected continuum in the
    IMACS search data,   implying rest-frame equivalent widths, $W_0 \ga 7-10$ \AA.  
     Therefore,  the combination of these two different survey methods is valid. 

The COSMOS \citep{Murayama}, and SDF (Kashikawa et al. 2011) samples 
are derived in similar ways. Both are selected from Subaru/SuprimeCam 
observations through the NB816 filter.    However, the color cuts used to select the 
LAEs are not the same:  Murayama et al.
 require $iz - NB816 > 0.7$\footnote{The weighted average 
 magnitude, $iz$, is commonly used to represent the continuum under the NB816 
 band, and is derived from $i$ and $z$ fluxes: $f_{iz} = 0.57 f_i + 0.43 f_z$
  \citep{Murayama, Takahashi}.}, while the SDF LAEs are selected with $i- NB816 > 1.5$.  
  Comparison of these catalogs shows only a small difference.  All of the galaxies 
  in the SDF catalog have $iz - NB816  > 0.67$, and only two have $iz-NB816<$  0.70.   
  Conversely, 83\% of the Murayama et al. (COSMOS) catalog meets the SDF color cut. 
   Since the suitability of one color cut over another is unknown, we make no
    corrections, but note that this level of systematic uncertainty is present in the data.

\begin{figure} 
\plotone{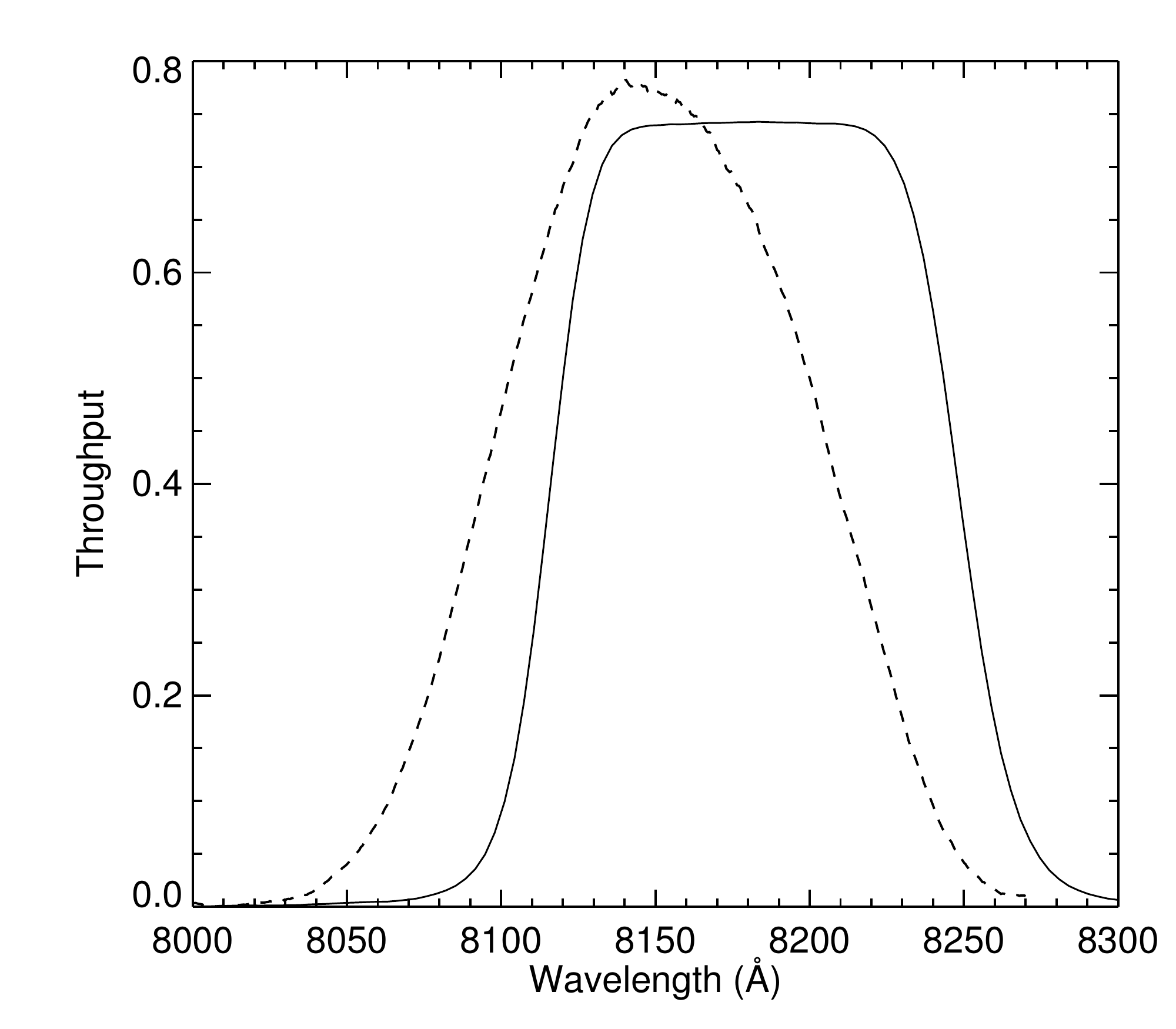} 
\caption{The NB816 throughput (dashed line)  is non-uniform, so the emission line flux 
inferred from NB816 magnitudes will, on average, be underestimated.  For comparison, we show
the much squarer transmission profile of the filter used in this work (solid line).} 
\label{filters} 
\end{figure}

  For the SDF sample,  
 we calculate the line fluxes and \lya\ luminosities, taking
  into account the non-uniform transmission of the NB816 
  filter (see Figure \ref{filters}).  These quantities are calculated from 
  the 2\arcsec\ aperture $z-$band and NB816 magnitudes, following the prescription in \cite{Kashikawa11}.   
  When redshifts are unknown, they are fixed to $z=5.70$ so that the line is assumed to fall at the center of the 
  filter bandpass.    A single aperture correction is applied to all galaxies in this sample, which we 
  estimate by  comparing SExtractor \citep{sextractor}  MAG\_AUTO  magnitudes to the 
  2\arcsec\ diameter magnitudes. This approach is preferable to simply adopting the
   MAG\_AUTO magnitudes, which are unreliable at low S/N.  From this comparison, 
   we find a median aperture correction of -0.38 magnitudes to the NB816 photometry, 
   indicating a +0.15 dex correction to the \lya\ luminosity.

 Considerable efforts to follow up the SDF sample have confirmed 46/89 LAE candidates and 
 identified four \oii\ emitters \citep{Shimasaku, Kashikawa11}.
 The spectroscopic confirmation  is 
 highly complete at the bright end, but falls off to fainter luminosities. 
Excluding the \oii\ emitters, and dividing the sample at  
$L(2\arcsec)=  10^{42.6}$  erg s$^{-1}$,  31/36  (86\%) of the
 brighter LAE candidates are confirmed, but only 15/39 (38\%) of the fainter
  candidates are confirmed.  Since narrowband imaging is 
  subject to the selection of spurious (no emission line) sources at faint 
 magnitudes, we restrict the sample to $L(2\arcsec) >  10^{42.6}$  erg s$^{-1}$ where
  the confirmation rate is high.  The five unconfirmed LAE candidates
 with these bright luminosities are included, as their unconfirmed status is likely the 
 result of slit-mask design.

We calculate the differential luminosity function for the SDF LAEs
by  determining the effective volume over which each object can be both 
detected with 
NB816 (2\arcsec) $<26$, and selected with $i-NB816 > 1.5$.  Within the
 725 arcmin$^2$ of the SDF, these volumes range from
  $1.7\times 10^{5}$ Mpc$^{3}$ for $L \sim 10^{42.6}$ erg s$^{-1}$  to 2.3 $\times 10^{5}$ Mpc$^{3}$ for $L>10^{43}$ erg s$^{-1}$.    For the five LAEs that are not spectroscopically confirmed we adopt an
   effective
volume defined by the redshift range that corresponds 
to the FWHM of the NB816 filter: 1.9 $\times 10^{5}$ Mpc$^{3}$.  
Next, a completeness correction for each object, as a function of its NB816 
magnitude (K. Shimasaku, private communication), is applied. 
  The differential luminosity function from the SDF is plotted in Figure \ref{lf_bins}.

 Combined with the SDF,  COSMOS is an excellent complement, covering nearly
  10 times the area to a shallower depth.   Over 1.86 degrees$^2$, 
\cite{Murayama} report 111 LAEs.   Spectroscopic redshifts for this sample are unavailable, 
so like the unconfirmed SDF LAEs, we adopt a volume spanned by the FWHM of the filter
 bandpass:  $1.9 \times 10^6$ Mpc$^{3}$. Likewise, line fluxes and luminosities are 
 calculated in the same way that was done for the SDF LAEs.    However, we note that,
  for the COSMOS sample, the best estimate of the total luminosities are taken from 3\arcsec\ diameter 
  apertures.  In Figure \ref{lf_bins}  we show the differential luminosity 
function for the 65 LAEs brighter than 10$^{43}$   erg s$^{-1}$.  Although the precise completeness 
of this sample is not known, comparison to the SDF LF shows that the COSMOS LAE catalog is 
complete above this conservative cut.   As we will show below, the slightly lower normalization of the 
COSMOS \lya\ LF is actually a shift in luminosity of about 0.05 dex, which results from the assumption
 that all of  these emission lines  are centered  in the NB816  bandpass.

\subsection{Maximum Likelihood Derivation} 
\label{lyalf_mle} 
So far, in our calculation of the \lya\ LF shown in Figure \ref{lf_bins},  we have assumed
 that the observed 
luminosity, $L$, is a good approximation to the true luminosity $L_0$.    However, in
 some cases, this approximation will fail. 
In the blind spectroscopic data, a luminous galaxy can be detected as a faint object if it 
falls outside of the slit.  
 Similarly, with narrowband imaging through the non-uniform transmission of the NB816 filter, 
 luminous galaxies can be observed with faint luminosities 
if the line is transmitted far from the center of the bandpass.  Not only are the luminosities altered by slit and filter transmission effects,
 but the volume from which a galaxy is drawn also becomes uncertain.  Relative to fainter galaxies,  
 intrinsically  bright line emission can be detected over a larger volume because these galaxies
  can be observed further from the slit- or bandpass- center.  In short, blind-spectroscopic observations 
  and narrowband imaging surveys are subject to the same problems: 
when precise slit-positions or redshifts are unknown, true luminosities are uncertain and volumes
 are not well-defined.

A maximum likelihood approach to the luminosity function parameter estimation (e.\ g.\ \citealt{STY}; STY) offers a 
natural solution to this problem,
as it allows the flexibility to model these effects of uncertain luminosity and volume.    With a Monte Carlo 
simulation (described below) we can 
determine the probability  $k (L_0 \rightarrow L)$ that a galaxy with some true luminosity, 
$L_0$, will be observed with some other luminosity $L$ \citep{Gronwall,Reddy08}.    The luminosity function
 that we observe as a function of $L$ is then a convolution of the true luminosity
  function with these kernels, $k (L_0 \rightarrow L)$:  
\begin{equation} 
\phi(L) = \int k (L_0 \rightarrow L) \phi(L_0) dL_0.
\label{transphi} 
\end{equation} 
Here, $\phi(L_0)$ represents the Schechter parameterization of the luminosity function: 
 $\phi(x)dx = \phi^*x^{\alpha} e^{-x}dx$ , where $x = L/L^*$.   In the limit where there is no
  scatter between $L_0$ and $L$,  $k(L_0 \rightarrow L)$ becomes a delta function, and
   Equation \ref{transphi} reduces to $\phi(L) =  \phi(L_0)$.   

This treatment can be applied directly to the STY maximum likelihood parameter estimation,
 where we choose $\alpha$ and $L^*$ that maximize the likelihood of observing the set of $N$ galaxies 
 in our sample, with luminosities $L_i$: 
\begin{equation} 
\displaystyle \mathcal{L}(\alpha, L^{*}) = \prod_{i=1}^{N}  \frac{ \phi (L_i) } {\int \phi(L) dL }.  
\label{maxlike_simple}
\end{equation} 
Including the convolution shown in Equation \ref{transphi}, the likelihood function becomes:   
\begin{equation}
 \displaystyle \mathcal{L}(\alpha, L^{*}) = \prod_{i=1}^{n}  \frac{  \int k (L_0 \rightarrow L_i) \phi(L_0) dL_0}
 {\int dL \int dL_0 k (L_0 \rightarrow L) \phi(L_0) } 
\label{maxlike_best}
\end{equation} 
Since $\phi^*$ appears in the numerator and denominator, the
 normalization is set by the 
number of galaxies that are observed.   To calculate the normalization,
 while accounting for the 
covariance between $\alpha$, $\phi^*$, and $L^{*}$, we calculate the 
number of galaxies, $N'$, 
predicted by each model:
\begin{equation} 
N' = \int dL \int dL_0 k (L_0 \rightarrow L) \phi(L_0) \times Volume. 
\label{npred} 
\end{equation} 
The appropriate survey volume here is the one that is matched to the 
simulation used to derive $k(L_0 \rightarrow L)$.   In the following 
calculations, the 
survey volume is not a function of luminosity, because galaxies are
 uniformly
 simulated out to some maximum distance from slit-center 
 (or wavelength from away
  from the bandpass-center).  Under this approach,
  $k(L_0 \rightarrow L)$ reflects the incompleteness to fainter
   galaxies  when their simulated positions/wavelengths
   render them undetectable. 
Finally, we include the (Poisson) probability of observing
 the $N$ galaxies in our survey, 
given the model prediction of $N'$:
 \begin{equation}
  \mathcal{L}(\alpha, L^{*}, \phi^*) =  \mathcal{L}(\alpha, L^{*}) * P(N | N') . 
  \label{lthree} 
 \end{equation} 
 The joint constraint from the three different 
 surveys that we use in this paper is simply the product of the
  likelihood functions, $ \mathcal{L}(\alpha, L^{*}, \phi^*)$, for each.

\subsection{Implementation of the maximum likelihood parameter estimation} 
\begin{figure*} 
\plotone{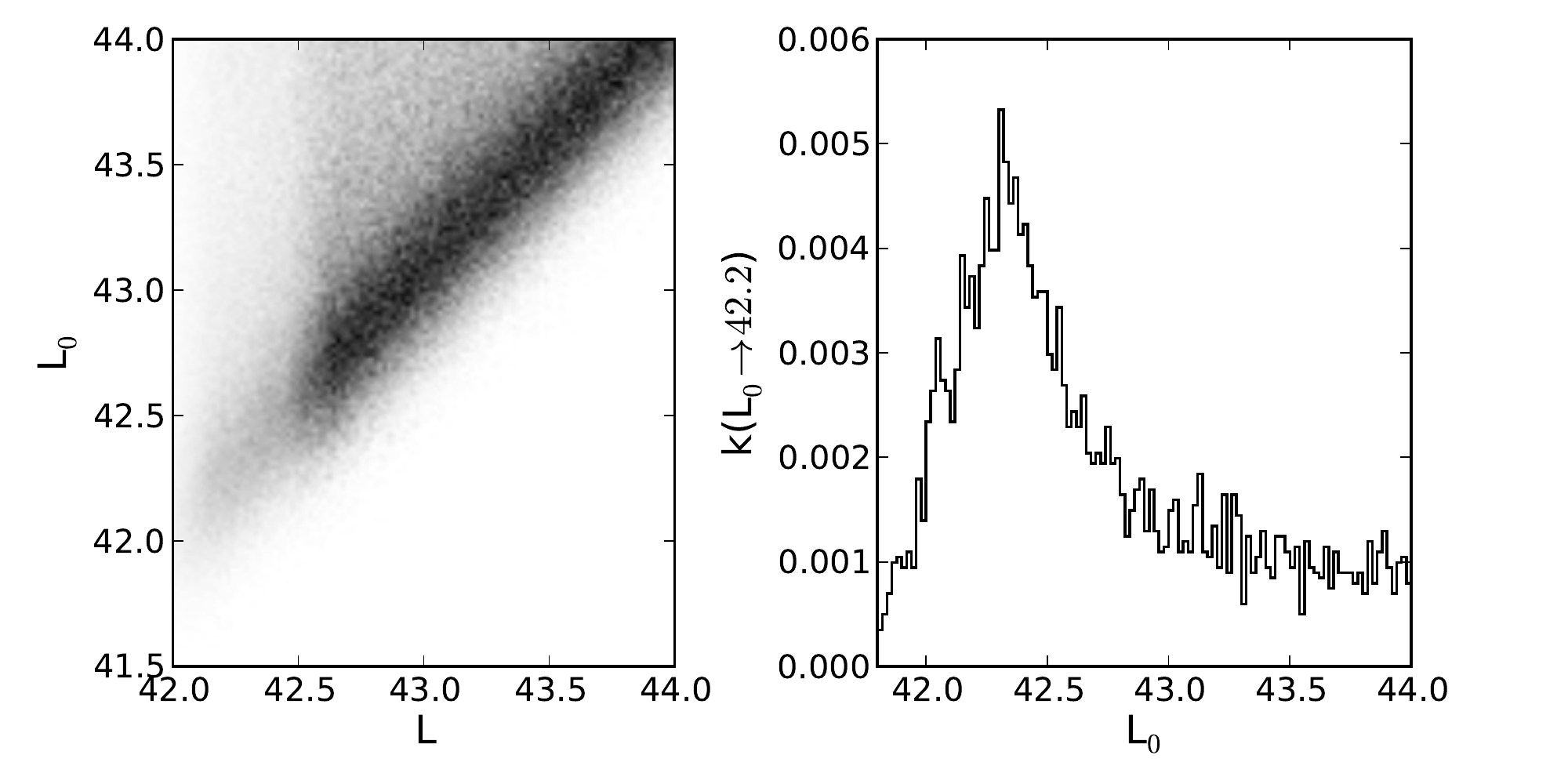}
\caption{ \textit{Left--}  The convolution kernels $k(L_0  \rightarrow L)$ describe the 
probability of a galaxy with luminosity $L_0$ being observed with some other 
luminosity $L$.  While these probabilities are strongly peaked on the diagonal, where $L_0 \sim L$, 
there is a non-trivial tail where galaxies with large $L_0$ are detected with faint $L$.  These are 
galaxies that 
fall partially outside of slits.   This example is for the 6 LAE case, so the redshift identification 
incompleteness is apparent at $L \la 10^{42.5}$ erg s$^{-1}$. \textit{Right--} A vertical slice of 
the left panel, taken for $L = 10^{42.2}$ erg s$^{-1}$ shows the fraction of galaxies that are
 detected with this luminosity but in actuality have other luminosities, $L_0$.  Again, the tail 
 to high $L_0$ comes from sources that fall partially outside the slit. }
\label{ptrans_fig} 
\end{figure*} 

Having outlined the steps needed to derive the best-fitting Schechter parameters, we next 
implement this method for both our spectroscopic data and bright-end data taken
 from the COSMOS and the SDF narrowband imaging surveys.   For each of these
  three surveys we use Monte Carlo simulations to calculate the convolution
   kernels $k(L_0 \rightarrow L)$.   With these quantities in hand, we
    find the $\alpha$, $L^*$ and $\phi^*$ that maximize Equation \ref{lthree}.

\subsubsection{Blind Spectroscopic Data} 
The quantities  that determine $k(L_0 \rightarrow L)$ for blind spectroscopy 
are: (1)  the photometric scatter, 
 and (2) the observed sizes of the \lya\ emission
in our 0\farcs5 seeing search data.   As mentioned in \S \ref{specfollow}, our 
DEIMOS observations, when compared to  the IMACS search data, indicate 
a photometric scatter with an RMS of 40-50\%, 
regardless of line flux.   The sizes of LAEs, on the other hand, require a more 
careful
consideration.      
Continuum observations
show that the majority of LBGs, including those with \lya\ emission, 
have compact sizes at $z\sim6$ \citep{Bouwens04, Ferguson, Henry10, Malhotra11, Gronwall11}.   Additionally, {\it Hubble Space Telescope} imaging of \lya\ emission from redshift 3-4 LAEs  supports the scenario where \lya\ emission is similarly
 compact \citep{Bond10, Finkelstein11}.   
However, two of the securely identified LAEs  are clearly resolved
 in the DEIMOS followup observations, so we consider 
the impact of size.    

Because our sample is too small at this time to adopt an empirically motivated size 
distribution, we consider three cases to illustrate 
the effect of LAE sizes.  We assume constant sizes of 0\farcs5, 0\farcs75, and 1\farcs0
 (observed FWHM), and derive LAE parameter constraints (by the method 
described below).     This test shows that, when larger sizes are adopted,  the most likely
 values of $\phi^*$, $L^*$, and $\alpha$ are
unchanged within the precision of our parameter search 
 ($\Delta$log($L^*$) = 0.01,  $\Delta$log($\phi^*$) = 0.01, and $\Delta\alpha$ = 0.05).  
  Only the likelihood contours are slightly altered when larger sizes are used.  Relative to
   the constraints obtained when  0.5\arcsec\ FWHM are adopted,  1\arcsec\ sizes result in an 
   extension of the likelihood contours to values brighter by 0.1 dex for  log($L^*$),  and
    steeper by 0.1 for $\alpha$ (68\% confidence). 
 For quantifying the uncertainties, we 
 adopt the intermediate case where all LAEs have observed FWHM = $0.75$\arcsec\ in the
  IMACS search data.

With these assumptions,  we determine $k(L_0 \rightarrow L)$  by a Monte Carlo simulation 
where LAEs are randomly placed inside and near the slits, and noise is added. We consider
 an intrinsic luminosity range of  $L_0 = 10^{41.5} - 10^{44}$ erg s$^{-1}$.  
  The lowest simulated luminosity is  fainter than the limit of our search, so this allows for 
  photometric scatter into our selection.  The high-luminosity cutoff is arbitrary, but is justified
   since the volume density of LAEs becomes 
vanishingly small at this luminosity \citep{Murayama}.    This high$-L$ cutoff determines
 the maximum distance from slit-center where LAEs are placed.  A galaxy with 
  $L_0 = 10^{44}$  erg s$^{-1}$ and a size of 0.75\arcsec\ (FWHM) can present detectable
   line flux even from a distance of 1.7\arcsec\ from the slit-center.   Therefore, for the blind 
   spectroscopy, the volume used in Equation 
\ref{npred} is 3.4$\times 10^{4}$ Mpc$^{3}$. 

Since $k(L_0 \rightarrow L)$ describes the probability of a galaxy with luminosity 
$L_0$ being observed in our survey,
it is appropriate to include survey completeness in the Monte Carlo simulation.   
 This takes two forms.  First, the detection incompleteness, which is derived in  
 Appendix \ref{comp_app}, is included.    Additionally, since we are deriving LFs from inferred counts,
   we include a  redshift identification  
incompleteness defined as the ratio of the confirmed LAEs (columns four and five) to the
 inferred LAEs  (columns eleven and twelve) in Table \ref{rates}.   The bin with zero 
 inferred LAEs is assigned a spectroscopic completeness of zero. 

 Figure \ref{ptrans_fig} shows  $k(L_0 \rightarrow L)$ with the spectroscopic
  incompleteness included for the six-LAE case.    
The left panel shows that, in this two
dimensional space, the probability is peaked on the diagonal because
 the most likely true luminosity, $L_0$, is the one that we observe, $L$.  
 However,  a tail of galaxies with high $L_0$ and low observed $L$ are 
 also present when 
galaxies fall partly outside of a slit.   On the left-hand side of this two 
dimensional space, at $L < 10^{42.5}$ erg s$^{-1}$, the spectroscopic 
incompleteness is apparent.    The right panel of Figure \ref{ptrans_fig} 
shows a vertical slice of the left panel to illustrate the fraction of galaxies
 at each $L_{0}$ that are detected with $L = 10^{42.2}$ erg s$^{-1}$.   Because of photometric scatter, 
 there is a significant probability of a galaxy having a fainter true
  luminosity, $L_0$, than is observed.  It is worth noting that, if a set of 
  Schechter parameters are assumed, we can use this slice in  Equation \ref{transphi} 
to get $\phi(L = 10^{42.2}~{\rm erg~s^{-1}})$.  

\begin{figure*} 
\plotone{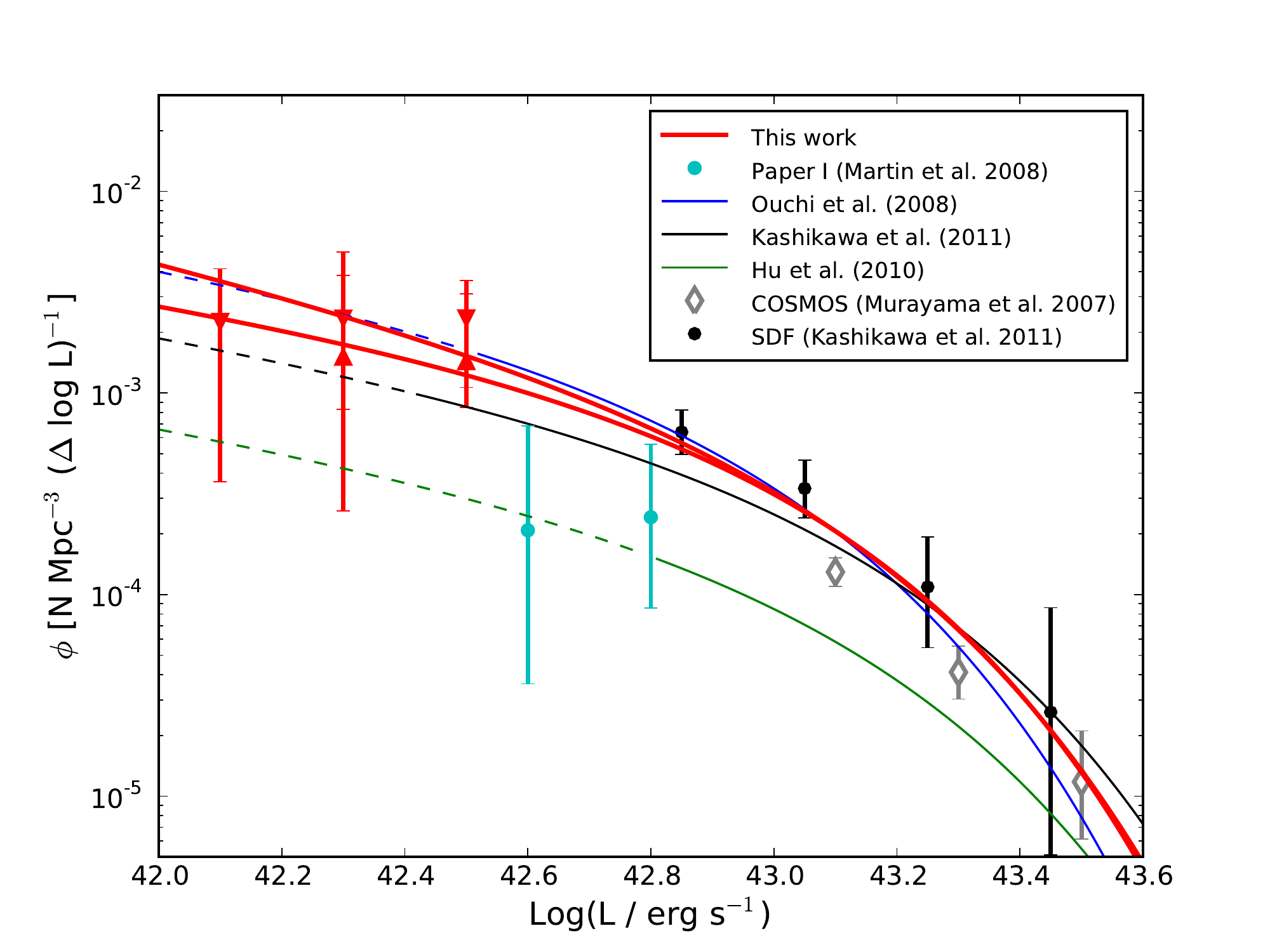} 
\caption{Best-fit LFs to the inferred LAE counts (high and low), shown in red, are
 compared to other derivations of the redshift 5.7 luminosity function.  Curves are dashed at 
  luminosities where they are extrapolated beyond the data in their various studies.  For
   the \cite{Ouchi08, Kashikawa11} and \cite{Hu10} studies we show the fits for $\alpha  \equiv -1.5$, 
   which is intermediate to the values favored by our data.} 
\label{lfs_with_fits}
\end{figure*}

\subsubsection{Bright-end Narrowband Imaging Data}    
  To calculate $k(L_0 \rightarrow L)$ for the narrowband imaging samples, we must take
   two scenarios into consideration: when the redshifts of the emission lines are known, 
   as they are for most (but not all) of the SDF LAEs, and when the redshifts are unknown, 
   as is the case for the COSMOS narrowband imaging sample.      When redshifts are unknown, 
  emission lines with some $L_0$ are placed randomly at wavelengths between 8050 \AA\ and  8250 \AA.   
    Therefore, the appropriate volumes 
  used in Equation \ref{npred} are $3.2 \times 10^{5}$ Mpc$^{3}$ for the SDF and 
    $3.0 \times 10^{6}$ Mpc$^{3}$ for COSMOS. 
  Additionally, a normally distributed flux error is added, with
   $\sigma =  2 ~{\rm or}~ 5 \times10^{-18}$ \flux\  for the SDF and COSMOS simulations.   
  
  As with the blind spectroscopic data, these Monte Carlo simulations include detection
   completeness.   While the COSMOS narrowband imaging sample is assumed to be 
   complete at $L > 10^{43}$ erg s$^{-1}$, the SDF data must be corrected.   Although 
  the incompleteness for this sample is a function of the NB816 magnitude, it varies
   slowly from 75 - 90\% over the range that we are interested.  Furthermore, the line
    luminosity forms a fairly tight correlation with the NB816 magnitude, with an RMS
     scatter of 0.1 dex. Therefore, at each luminosity we adopt 
 a completeness appropriate for the mean NB816 magnitude of galaxies with 
 that luminosity. 
 
 When the redshifts of the emission lines  are known, a different set of 
 kernels is used in Equation \ref{maxlike_best}. 
 In this case, the simulation differs from the case of unknown redshifts, 
 because the observed $L$ have been corrected for the transmission
  through
 the filter curve shown in Figure \ref{filters}.   Therefore, in this simulation there
  is no scatter between $L_0$ and $L$ from the unknown filter throughput.  
 In the absence of noise, these kernels would be a set of $\delta$-functions.   
 However, as with the case of unknown redshifts, noise and detection 
 incompleteness are included.    For the purposes of determining the normalization, 
 we adopt the scenario used for unknown redshifts in Equation \ref{npred} for all SDF LAEs.

\subsection{Results: Luminosity Function Constraints} 
With the various $k(L_0 \rightarrow L)$ in hand,  we evaluate Equation \ref{lthree} for each
 survey, over a grid of $\alpha$, $L^{*}$, and $\phi^{*}$.   
The final likelihood function is given by the product of the individual likelihood functions, 
$\mathcal{L}(\alpha, L^{*}, \phi^*)$, for each survey.   
The parameters that maximize this quantity are given in Table \ref{bestfits}, for both the 
3-LAE and 6-LAE LFs.   

\begin{figure*}
 \plotone{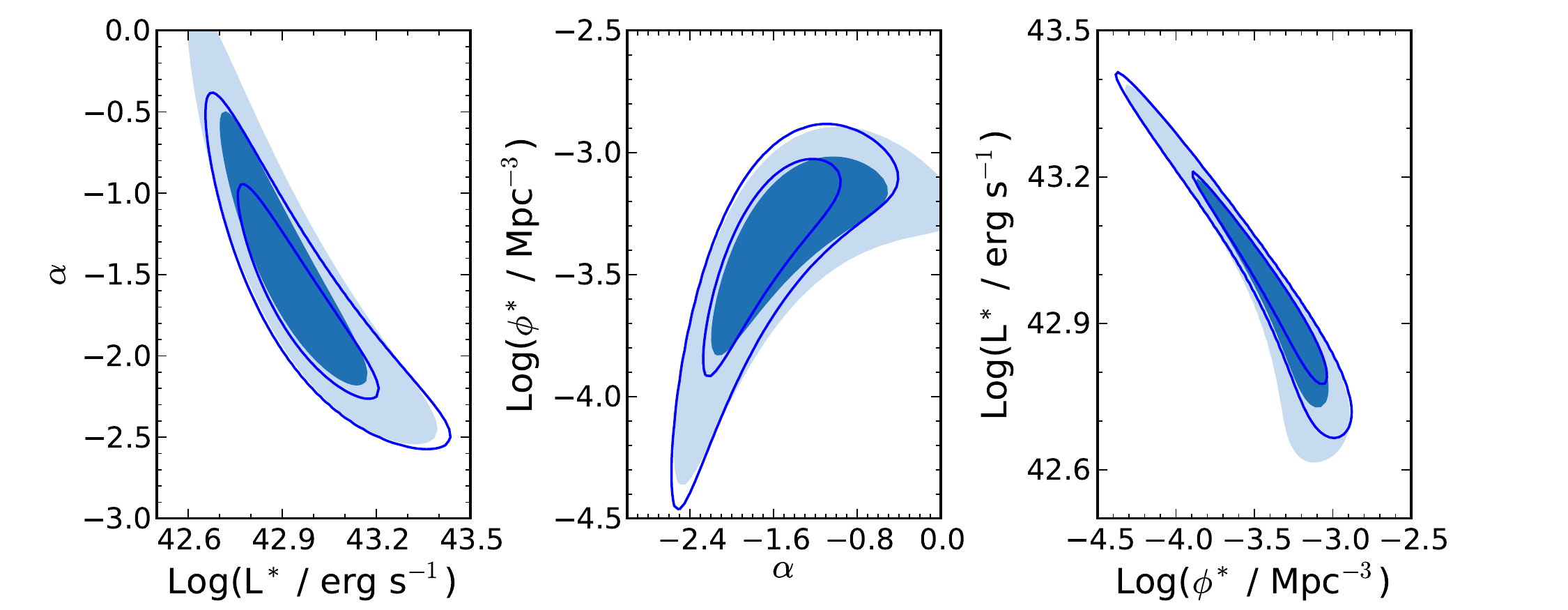} 
 \caption{Likelihood contours show the Schechter parameter constraints from the 
 Multislit Narrowband Spectroscopic sample (this paper), combined with the bright-end narrowband 
 imaging data from COSMOS and the SDF.  The 68 and 95\% confidence 
 intervals are represented by the shaded contours for the 3 LAE LF and by the open curves for the 6 LAE LF.    }
 \label{contours}
\end{figure*} 

\begin{deluxetable*} {cccc}
\tablecolumns{4}
\tablecaption{Most Likely Schechter Function Parameters}  
\tablehead{
\colhead{Sample} & \colhead{log($\phi^* / {\rm Mpc}^{-3}$)} & \colhead{log($L^*/{\rm erg s}^{-1}$) }  & \colhead{ $\alpha$  } 
}
\startdata
3 LAE &    $-3.29^{+0.27}_{-0.52}$ & $42.91^{+0.19}_{-0.21}$ & $-1.45^{+0.92}_{-0.70}$ \\
6 LAE &   $-3.35^{+0.31}_{-0.56}$ &  $42.96^{+0.24}_{-0.16}$   &  $-1.70^{+0.73}_{-0.57}$ 
\enddata
\label{bestfits} 
\end{deluxetable*} 

The best-fitting Schechter-functions are plotted in Figure \ref{lfs_with_fits}.   One feature of maximum likelihood
 parameter estimation is that the 
most likely model does not always run through the data points.  This is indeed the case in Figure \ref{lfs_with_fits}.  
The discrete points describing
 the COSMOS LAEs are derived by assuming that  all objects are observed through the peak of the 
NB816 throughput curve.   Consequently, on average, the true luminosities, $L_0$, of these
 LAEs will be brighter than what we have observed.     Since we model these effects with our
  maximum likelihood estimation, we expect the Schechter functions plotted in Figure
   \ref{lfs_with_fits} to fall above 
the discrete data points for the COSMOS narrowband imaging data.  On the other hand, 
since we have taken the bandpass into account for the 85\% of the SDF LAEs that are spectroscopically 
confirmed, we expect the most likely Schechter function to closely follow the discrete SDF points in Figure \ref{lfs_with_fits}. 
In fact, if we exclude the SDF data from our fit, the bright-end of the best-fit Schechter function 
   falls close to the same location, approximately 0.05 dex brighter than the COSMOS  narrowband
    imaging points shown in Figures  \ref{lf_bins} and \ref{lfs_with_fits}.  Correcting these points in 
    Figures \ref{lf_bins} and  \ref{lfs_with_fits} for this average attenuation, would bring them
     in line with the SDF data and the Schechter function fit.

Figure \ref{contours} shows the 68\% and 95\% confidence intervals for these parameter constraints.  
 Clearly, owing to our small sample of faint LAEs, a large  range of acceptable faint-end slopes can
  describe our data.     However, as our spectroscopic followup becomes more complete, 
the luminosity function parameters will be more tightly constrained.

\section{Improving the faint-end slope constraint with more followup spectroscopy} 
\label{improve}
It is not surprising that the luminosity function parameter constraints which we show in 
Figure \ref{contours}   allow a wide range of Schechter parameters. 
 At this time the faint-end of our LF is based on only a few confirmed LAEs.    Our continued 
 efforts at followup spectroscopy will narrow the allowed parameter space by increasing 
 the  sample size.  Furthermore, the addition of our 15H field, when spectroscopy is complete, 
 will approximately double the number of LAEs over what is present in the COSMOS field. 

\begin{figure*} 
\plotone{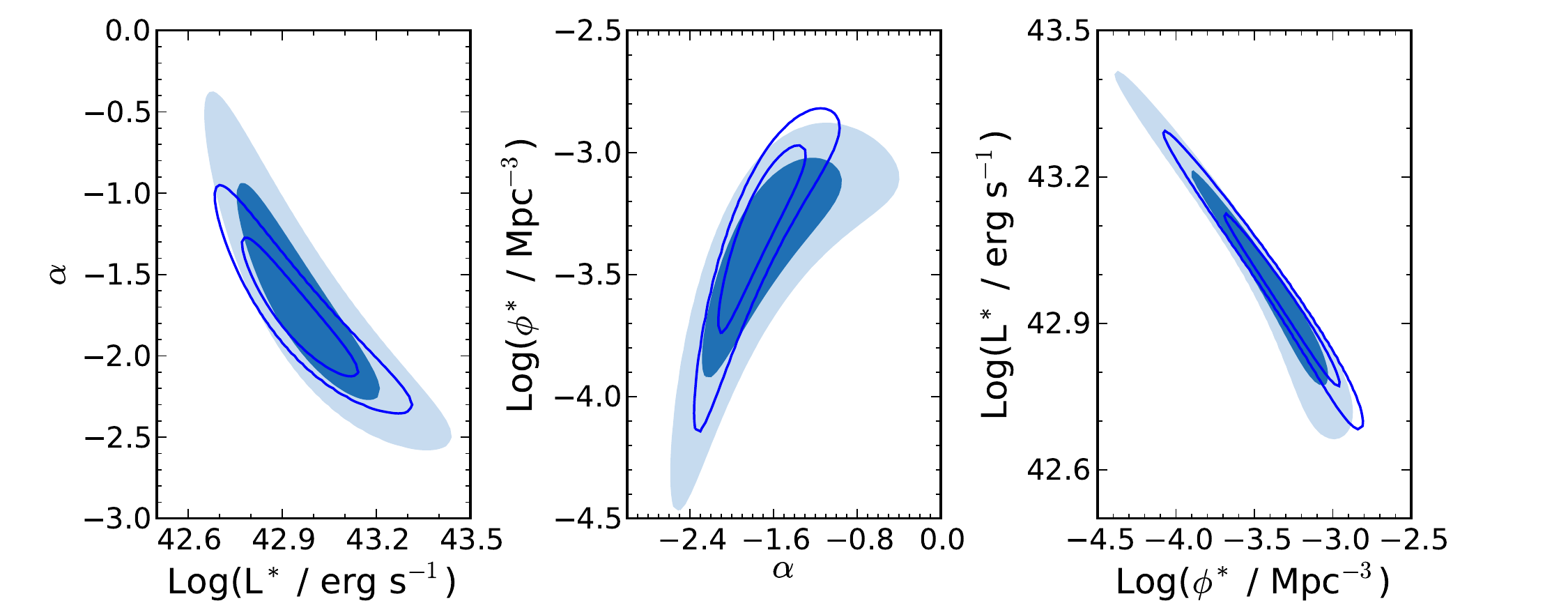} 
\caption{Likelihood contours show the  Schechter parameter constraints that we can achieve 
when the spectroscopic followup is 90\% complete in both of our survey fields (curves).  An improvement relative to our current constraints (shaded) is 
expected.   For simplicity, we compare only the LF
 derived from 6 LAEs to a simulated sample 
derived from the 6 LAE LF Schechter parameters.   The bright end data are unchanged from 
what we have used in this study. Contours are shown for 68 and 95\% confidence.    } 
\label{contours_future1} 
\end{figure*}

It is therefore of interest to address how strict the constraints from our survey may ultimately prove.
  To explore this,  we create a mock catalog of galaxies
   with $\phi^* = 10^{-3.35}~ {\rm Mpc}^{-3}$,  $L^* = 10^{42.96} ~{\rm erg~ s}^{-1}$, 
   and $\alpha= -1.70$ (our 6 LAE LF in Table \ref{bestfits} and the  contour
    curves in Figure \ref{contours}).   These galaxies are distributed throughout
     the volume of {\it both} our COSMOS and 15H  IMACS fields, out to 1.7\arcsec\ from slit-center, 
     similar to the Monte-Carlo simulations in \S  \ref{lyalf_mle}.    Slit-losses, noise, and detection
      completeness are included. Finally, we assume a spectroscopic completeness of 90\% at all 
      luminosities, since achieving 100\% completeness is usually impractical.   The resulting sample of
       LAEs in both fields is approximately 40 galaxies. 
After regenerating the convolution kernels to include the appropriate spectroscopic completeness, we 
calculate the likelihood function, $ \mathcal{L}(\alpha, L^{*}, \phi^*)$  over the grid of $\alpha, L^{*}, \phi^*$ 
(leaving the bright-end data unchanged).  Contours shown in Figure \ref{contours_future1} illustrate the 
improvements in the constraints that we can expect 
to find from completing our spectroscopic followup.     For the increased sample size in this simulation, 
the 68\% confidence interval on 
the faint-end slope is reduced by 50\%.  Likewise, under our 
current constraints, the 
space density of LAEs at $L = 10^{42}$  erg s$^{-1}$  spans a
 factor of 10 (68\% confidence), but should be decreased to
  less than a factor of three when both fields are 90\% complete. 

\section{Discussion} 

\subsection{Comparison to previous measurements of the redshift 5.7 \lya\ LF} 
In Figure \ref{lfs_with_fits} we compare the LFs derived in this work to those presented in 
previous studies at redshift 5.7.   A number 
of items are readily evident.  First,  the shape ($\phi^*$ and $L^*$) of the LF that we derive 
from the SDF data is different from the shape obtained by \cite{Kashikawa11}, {\it from the same bright-end data}.   
This  variation is not surprising, given our different analysis.   Most notably, we have 
adopted luminosities calculated 
from the NB816 and $z$-band magnitudes, whereas Kashikawa et al.\ use luminosities 
derived from their followup spectroscopy.   Given the uncertainties associated with estimating
 slit-losses,  imaging-based measurements may be a better approximation to the total \lya\ luminosity-- provided 
 that aperture magnitudes are corrected to total magnitudes, and the redshifts are used to calculate the true
  luminosity observed through the NB816 filter.

Figure \ref{lfs_with_fits} also compares our bright-end LF derivation with the similar redshift 5.7 \lya\ LF from
 the Subaru/XMM-Newton Deep Survey 
(SXDS; \citealt{Ouchi08}).   While the agreement between the plotted Schechter functions is good, this may be 
fortuitous. 
Limited spectroscopic followup  in the SXDS could allow significant contamination, since narrowband imaging can 
sometimes 
select LAE candidates which have no line emission at all.   This point is illustrated clearly in  the  color magnitude
 plots used to select LAE candidates  (Figure 7 in \citealt{Ouchi08}; Figure 1 in \citealt{Murayama}; Figure 7 in
  \citealt{Shimasaku}).  When the limit of the NB816 photometry is approached, there is significant scatter about the 
  threshold that defines an emission line source.  While the use of optical ``veto'' bands will eliminate some
   spurious emission line candidates, even this method  
could break down for faint galaxies,  where modestly red galaxy SEDs may be undetected at bluer wavelengths. 
This contamination may account for the LAEs which \cite{Kashikawa11} do not detect in followup spectroscopy.  
Likewise, the faintest
{\it confirmed} LAEs in the SXDS \citep{Ouchi08} are about one magnitude brighter in NB816 than the limit of that sample.    
More spectroscopic followup will be needed 
to determine the reliability of the fainter \lya\ candidates selected from narrowband imaging. Such observations would 
 provide better constraints on 
the space density of LAEs at luminosities around the knee of the LF, and reduce the covariances shown in Figures
 \ref{contours} and \ref{contours_future1}. 

Two other measurements deviate from the LF that we derive here:  \cite{Hu10}, and \clm.  
  These bright and intermediate luminosity data would require a discontinuity from our new faint-end results.    A 
  Schechter function can not be fit  which simultaneously goes through these data and our new data.  Therefore,
   it is clear that both the \clm\ and Hu et al.  samples are
incomplete.       Indeed,  incompleteness in the Hu et al.\ result is discussed in \cite{Kashikawa11}.  Because 
Hu et al.\ require spectroscopic confirmation, but use shorter observations than Kashikawa et al., it is argued that the 
incompleteness is simply a failure to re-detect some real LAEs.  On the other hand, for our 
pilot survey, comparison of our current completeness to the detection completeness derived in \clm\ suggests that 
the prior estimate was overly optimistic.   For the current data, near 100\% completeness is reached for detections 
of better than 10$\sigma$ significance.    Shifting the completeness curve for the present data 
(Appendix \ref{comp_app}) by a factor of four to five in flux allows us to revise 
the completeness estimate used for the shallower data in \clm.    This exercise suggests that our 
previous LF was 30-70\% complete, rather than near 100\%, as was assumed.    Revising these 
completeness corrections would bring the \clm\ points more closely in line with   our new LF and the \cite{Ouchi08} result.

In \dressler, our analysis of the foreground populations suggested that the remaining 
counts of LAEs followed a similar faint-end slope, with $\alpha \sim -2$.  With our 
followup spectroscopy, we now find that the most  likely slope is  slightly shallower,
 although $\alpha = -2$ is well within the 1$\sigma$ 
measurement uncertainty.  Nevertheless, we identify the source of this difference
 (for the COSMOS field) in Appendix \ref{fgsec}.  Briefly, the foreground emission
  line galaxies that we have identified so far in our followup add up to the counts
   predicted by the foreground model adopted in \dressler.  Since we 
expect that a significant fraction of the unidentified emission lines listed in 
Table \ref{counts1} are also foreground objects, we expect these counts to be
higher than the most likely model adopted in \dressler.    Because of this, the
 fraction of LAEs among our detected emission lines will be smaller than
we previously estimated in \dressler.    As we show in Appendix  \ref{fgsec}, 
the most likely cause of this underestimation is a strong presence of \hb\ emitters 
from an overdensity at $z\sim0.68$ in the COSMOS field.

Finally, we note that the cosmic variance for our faint LAE sample, 
at this time, makes a negligible contribution to the error budget because 
the Poisson errors on our small sample are much larger than the cosmic variance.    
On the other hand, at the bright end, the excellent agreement between the 
narrowband imaging samples from COSMOS,  the SDF, and the  SXDS  is better than 
the approximately 20-30\% cosmic variance that we expect for these surveys \citep{TS09}.

\subsection{Comparison to faint \lya\ LFs at lower redshift} 
\begin{figure*} 
\plotone{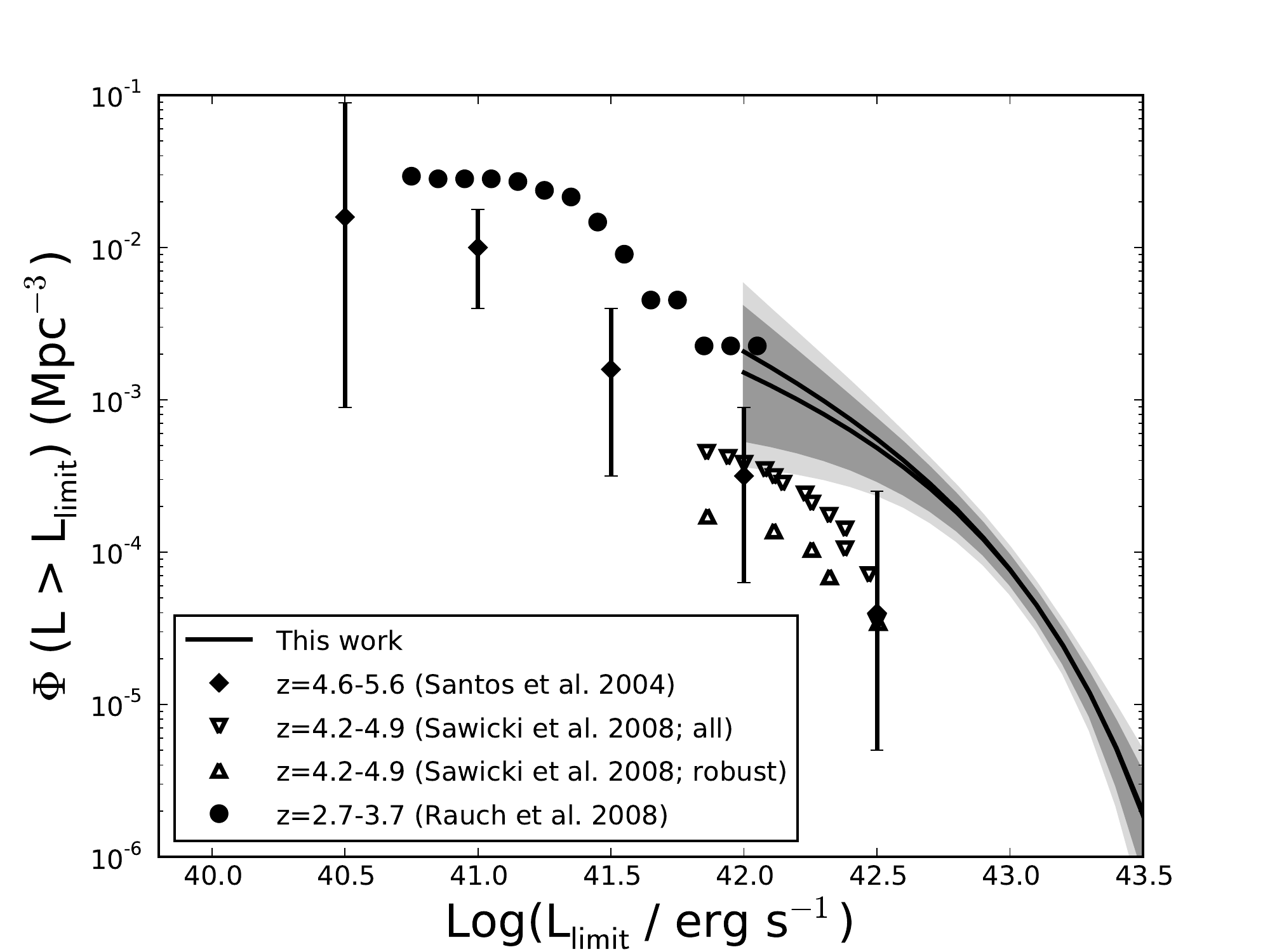} 
\caption{Spectroscopic surveys measure the cumulative  \lya\ luminosity functions to very faint luminosities at $z\ga3$.  
These include a search for strongly lensed LAEs  
\citep{Santos}, a serendipitous survey \citep{Sawicki08}, and a very long exposure with the Very Large
 Telescope \citep{Rauch08}.  The shaded
area show the 68 and 95\% confidence intervals from our LF fits, where, for simplicity we have taken the
 lower bounds from the 3 LAE LF and the upper bounds from the 6 LAE LF.    } 
\label{cf_faint} 
\end{figure*}

So far, we have compared our \lya\ LF to previously measured
 \lya\ LFs at redshift 5.7.  By necessity, this discussion focussed on the bright-end
data, since no faint-end measurements existed prior to the one which 
we present here.   In this section, we turn to lower-redshift spectroscopic 
surveys which measure LAEs to very faint luminosities at $z\sim3-5$.    
In Figure \ref{cf_faint}, we show three measurements of the cumulative \lya\ luminosity function.    
Each of these searches use broadband spectroscopy to detect \lya, either with
 lensing \citep{Santos},  through serendipitous 
detections \citep{Sawicki08}, or by exceptionally long exposures with the 
Very Large Telescope \citep{Rauch08}.

Figure \ref{cf_faint} shows that the space density of LAEs spans a factor of 
several between these different measurements.  Large statistical errors, 
coupled with differing analyses and a wide range of redshifts make it difficult to
 draw any conclusions.    The highest space density of LAEs is found
by \cite{Rauch08}, who also have the greatest sensitivity to low surface brightness emission.   
This could mean that they measure brighter luminosities
by including more low surface brightness flux, and also that they are able to detect extended,
 luminous LAEs that other surveys would miss.     
However, where Rauch et al.\ overlap with our survey, the agreement is good.   Lower 
densities are found at $z\sim4-5$ by  \cite{Santos} and \cite{Sawicki08}.     
At $L = 10^{42.5} $ erg s$^{-1}$ , where they overlap with the wider, shallower 
search from \cite{Ouchi08}, the density of LAEs found by Santos et al. and 
Sawicki et al. is several times lower.     The origin of this discrepancy is unclear.  
In short,  statistical uncertainties make it difficult to 
determine if the faint end of the \lya\ LF evolves at $z\sim3-6$.   Larger 
spectroscopic samples of faint LAEs are needed.

\subsection{Comparison to UV-selected $i-$dropout samples} 
\label{lfsim_sec}
Surveys for LAEs have have often been motivated by the idea that line emission 
can be more easily detected than 
continuum when galaxies become very faint.   This rationale is especially true  
for our spectroscopic search, which includes lines to fainter
luminosities than  most other \lya\  searches.   For example, at the median LAE 
luminosity in Table \ref{LAEdata} ($L = 10^{42.4}$ erg s$^{-1}$), a galaxy with 
rest \lya\ equivalent width\footnote{Here, and in the discussion that follows, we 
assume that  $f_{\lambda} \propto \lambda^{\beta}$ with $\beta = -2.0$.   While 
some evidence exists for bluer slopes at the highest redshifts \citep{Bouwens09, Bouwens10_slopes}, 
these claims 
are disputed (\citealt{Dunlop}; see also, \citealt{Finkelstein10}).
  Until more conclusive evidence for bluer slopes is found, we adopt a flat UV-slope when 
  converting from rest-frame 1350 \AA to 1216 \AA.}, 
   $W_0$,  larger than 190 \AA\ will have $M_{UV} > -18$.
At $z\sim6$, this luminosity is  beyond the limit of the $i-$dropout samples 
used to measure the UV-luminosty function \citep{Bouwens07}.   Likewise, at  the $L = 10^{42.0}$ erg s$^{-1}$  
limit of our survey,
a galaxy would need to have a rest equivalent width smaller than about 80 \AA\ to have a continuum
 flux-density which is bright enough for inclusion 
in the $i-$dropout sample. 

Of course, assessing whether LAE surveys are finding a significant population of previously
 unobserved galaxies is not straightforward.   The \lya\ equivalent width of star-forming galaxies
  displays a wide range of values, with a tail to quantities larger than
    $100-200$ \AA\ \citep{Shapley03, Shimasaku, Gronwall, Stark10, Stark11}.   
Consequently, a faint LAE could be the result of a low equivalent width 
emission from a UV-bright galaxy, or conversely, high equivalent width
 emission from an extremely faint galaxy.   No obvious trends exist
  between \lya\ luminosity and equivalent width (see \citealt{Henry10}), 
   although followup 
of UV-continuum selected $i-$dropouts shows that fainter continuum 
is correlated with stronger and more common \lya\ emission \citep{Stark10, Stark11}.    
Ultimately, since \lya\ equivalent width exhibits such a wide range of values, 
a statistical comparison with the ensemble of $i-$ dropout galaxies is most relevant.  

\begin{figure} 
\plotone{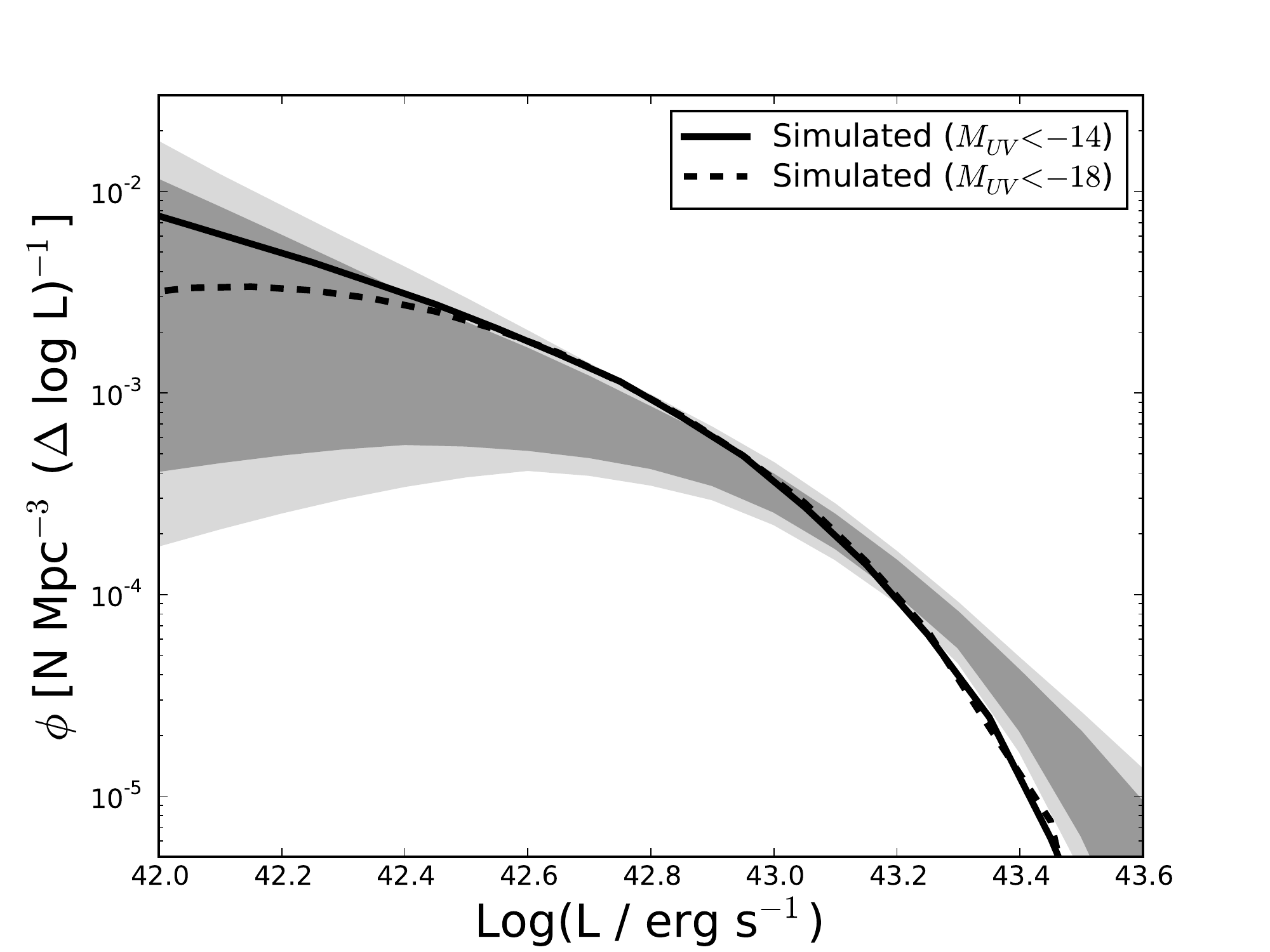} 
\caption{Simulated \lya\ LFs (solid and dashed curves) are compared to our LF constraints.  
 The shaded regions show our  68 and 95\% confidence
limits.  For simplicity, the lower bounds are taken from the 3 LAE LF and the upper bounds are
 taken from the 6 LAE LF.   The simulated LFs are derived by including galaxies to very 
 faint limits ($M_{UV} = -14$; solid), and also by limiting the simulation to luminosities 
 that have been observed in the UDF and UDF-Parallels ($M_{UV} < -18$; dashed). 
  The latter case turns over, representing an incompleteness to faint LAEs among the $i-$dropout
   samples.     At the limit of our MNS survey (10$^{42}$ erg s$^{-1}$), 50\% of LAEs have UV 
   luminosities fainter than the limit of current $i-$dropout samples.  } 
\label{lfsims} 
\end{figure}

To determine whether the LAEs in our sample are likely to be detectable as 
faint $i-$ dropouts, we generate a mock catalog of galaxies.  
Beginning with the UV-luminosity function of $z\sim6$ $i-$ dropouts 
(\citealt{Bouwens07}; $\phi^* = 1.4 \times 10^{-3}$ Mpc$^{-3}$, $M_{UV}^* = -20.24$, and $\alpha = -1.74$), 
we generate a sample of galaxies in a fixed volume, with UV-luminosities extending 
to $M_{UV} = -14.0$ (beyond the depth required for all LAEs in our survey to be
 detected in the continuum).  A rest-frame \lya\ equivalent width ($W_0$)  is assigned to each galaxy
  in this volume, drawing from an exponential distribution with a maximum of $W_0 = 2000$ \AA. 
Motivated by the observations shown in Stark et al.\ (2011; Figure 2), we adopt e-folding, 
or ``characteristic''  equivalent widths of 50 \AA\  for  $M_{UV} > M_{UV}^*$, 
and 20 \AA\ for $M_{UV} < M_{UV}^*$.  These scale factors approximately 
reproduce the distribution of rest equivalent widths and the fraction of sources
 with equivalent width less than 25 \AA.    Since narrowband imaging surveys
  cannot detect LAEs with $W_{0} < 25$ \AA, galaxies
which are given these low equivalent widths under our 
exponential parameterization are excluded from the \lya\ LF.  (Although we note that the \lya\ LF is not 
strongly dependent upon this cutoff.)  With each galaxy assigned an $M_{UV}$ and $W_0$, we calculate 
the \lya\ luminosity of each $i-$dropout.      To understand the impact of the current observational limit, we 
repeat this calculation including only galaxies brighter than  $M_{UV} = -18$.      The resulting simulated
 LFs are plotted in Figure \ref{lfsims}, where they are compared to our measured constraints on the \lya\ LF.   

The simulated LFs shown in Figure \ref{lfsims} illustrate the importance of 
surveys for faint LAEs.   With the equivalent width distributions chosen
 above, the \lya\ LF simulated  with only galaxies having $M_{UV} < -18$ (dashed curve) shows 
 a significant deficit of  LAEs relative to when
 the complete range of luminosities are included (solid curve).    At the $L=10^{42}$ erg s$^{-1}$ 
 limit of our survey, approximately 50\% of LAEs should have 
 $M_{UV} > -18$.   Galaxies as faint as $M_{UV} \sim -16.5$ are needed to bring these 
 two simulated LFs into
 agreement.    {\it  This result clearly indicates that we have uncovered a population of 
 galaxies whose continua are too faint to have been observed in the deepest space-based 
 dropout surveys.}    On the other hand, the narrowband imaging surveys
  for brighter LAEs are not uncovering a previously unobserved population.   
 Figure \ref{lfsims} shows that, brighter than  $L=10^{42.5}$ erg s$^{-1}$, the \lya\ luminosity 
 function is dominated by galaxies with $M_{UV} < -18.0$.

Figure \ref{lfsims} shows that the simulated \lya\ luminosity function falls short of 
observations at the bright end. 
A scale factor of 30 \AA\ rather than 20 \AA\ for the bright ($M_{UV} < M_{UV}^*$) $i-$ dropouts 
would bring the simulated luminosity function 
within the 68\% confidence interval shown in Figure \ref{lfsims}.   A larger scale factor is not
 implausible, as the sample of bright $i-$dropouts
targeted for spectroscopy by \cite{Stark11} is not large enough to derive very precise 
constraints on the luminosity dependence of the 
\lya\ equivalent width distribution.   Additionally,  it should be noted that the equivalent width distributions
measured by \cite{Stark11} are for galaxies that are {\it selected} as $i-$dropouts.   The selection function for
these galaxies is known to depend on \lya\ equivalent widths \citep{Bouwens06, Stanway07}, so the ``raw'' 
equivalent width distributions presented by Stark et al. may not strictly represent the parent population of star-forming galaxies. 


 One way to interpret the (incomplete) simulated \lya\ luminosity function of 
 galaxies 
 with $M_{UV} < 18$ is as a prediction of the \lya\ LF that one could measure from
  spectroscopic followup of $i-$dropout galaxies \citep{Jiang}.    However, correcting this LF
   to the ``true'' \lya\ LF of all galaxies is
 not possible.    We could speculate that the characteristic equivalent widths may be larger 
 for galaxies with $M_{UV} > -18$ (following the trend of more prevalent \lya\ emission among
  galaxies with fainter UV continuum luminosities). 
 Under this scenario, the discrepancy between the simulated LFs shown in Figure \ref{lfsims} would
  be amplified, because the faint-end slope of the ``true'' LF (simulated to a realistic low-luminosity cutoff) 
  would be even steeper.    Since we cannot know the equivalent width distribution of the faint, 
  continuum-undetected galaxies,   only dedicated emission line surveys will be able to measure 
  the  faint-end slope of the \lya\ LF.

\subsection{\lya\ line profiles} 
\label{lineprof_sec}
The transmission of \lya\ photons is sensitive to the ionization state 
of the IGM, so the shape of the \lya\ emission line profile may
offer a measure of the changing neutral hydrogen fraction. 
In a more neutral IGM,  the Gunn-Peterson \lya\ damping-wing will
 lower the peak flux of the line, effectively broadening the emission 
 and removing its asymmetric shape \citep{Ouchi10, Dayal}. 
  
\begin{figure} 
\plotone{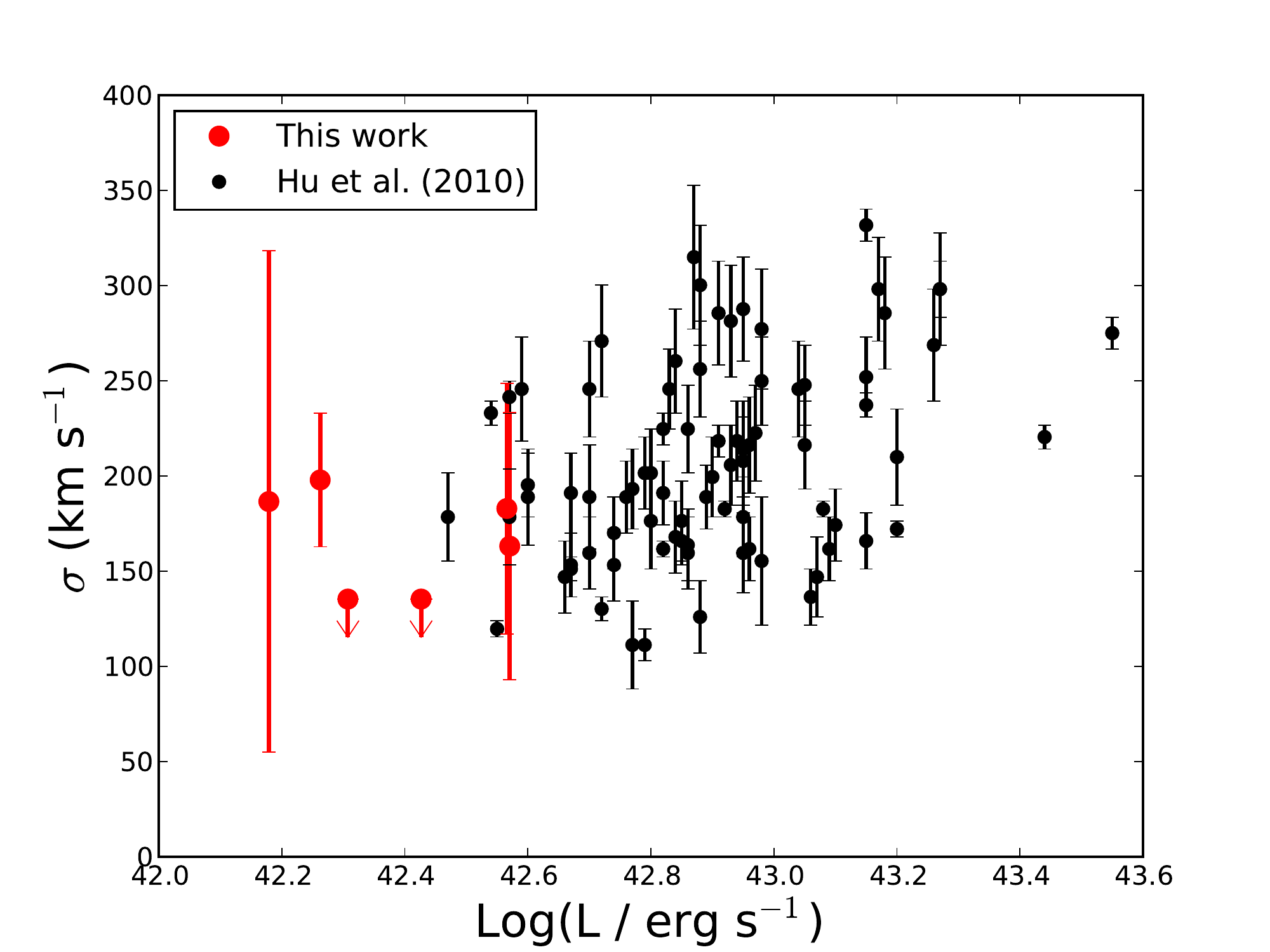} 
\caption{The \lya\ line width shows a positive correlation with luminosity.  
Here, $\sigma$ is determined in the same way for all LAEs: a truncated Gaussian is 
convolved with the instrumental resolution and fit to the observed line profile.  Two 
objects have line profiles that are consistent with being unresolved, and for these we 
plot upper limits that correspond to the instrumental resolution. } 
\label{lineprof_fig} 
\end{figure} 

Observational efforts have begun to search for a difference in \lya\ profiles 
observed at redshifts 5.7 and  6.5 \citep{Ouchi10, Hu10, Kashikawa11}, 
but the average line width is not significantly different between the two 
redshift bins. 
  However,  Hu et al. show that the
\lya\ line width displays a positive correlation with luminosity.  In 
Figure \ref{lineprof_fig} we combine our faint LAEs with the Hu et al. 
redshift 5.7 sample to extend the correlation to lower luminosities.  As expected from an 
extrapolation of the brighter sources, our newly measured line widths 
are among the narrower values.  The significance of the relation is improved: under the 
Spearman rank correlation test, the probability of the null hypothesis (no correlation) is 
decreased from  $1.6\times 10^{-3}$ to  $2.6\times 10^{-4}$ when our six faint LAEs are
 included. These probabilities, interpreted in the context of a Gaussian probability density 
 function, increase the significance of the correlation from 3$\sigma$ to better than 3.5$\sigma$. 

The luminosity dependence of the \lya\ line-width has 
important implications for how the effects of reionization on the line 
profile are interpreted.  
Comparisons made at different redshifts should also be made at
 fixed luminosity.   This luminosity dependence has not been accounted for 
when entire samples of redshift 5.7 and 6.5 LAEs have been
 stacked \citep{Ouchi10, Kashikawa11}. 
In a flux-limited sample, the redshift 6.5 LAEs will be, on 
average, more luminous.  
This could explain the observation that the redshift 6.5 line
 profiles are marginally broader than those at redshift 5.7.

The origins of the observed luminosity 
dependence are unclear.   One explanation offered by \cite{Hu10} is that 
the more luminous LAEs may simply be 
more massive, with intrinsically broader \lya\ lines.   Another possibility lies 
in the kinematics and conditions of the circumgalactic
medium.    While typical star-forming galaxies have redshifted \lya\ emission and 
blueshifted P-Cygni absorption \citep{Shapley03, Steidel10}, rare cases are seen
 where blueshifted material emits rather than absorbs \lya\ \citep{Erb10, McLinden}.  
 In these cases, the \lya\ emission is both luminous, and broader than average.   While
  Gunn-Peterson absorption will destroy blueshifted emission at redshift 5.7, the relative
   strength of the emission component at zero velocity may still induce a correlation between 
   line width and luminosity.  Measurements of LAE masses and systemic redshifts can distinguish 
   these scenarios,  but for reionization era galaxies, the required observations will remain 
   challenging until the {\it James Webb Space Telescope} operates.

\subsection{The contribution of faint LAEs to the hydrogen-ionizing background}
\label{reion_sec}

\begin{figure} 
\plotone{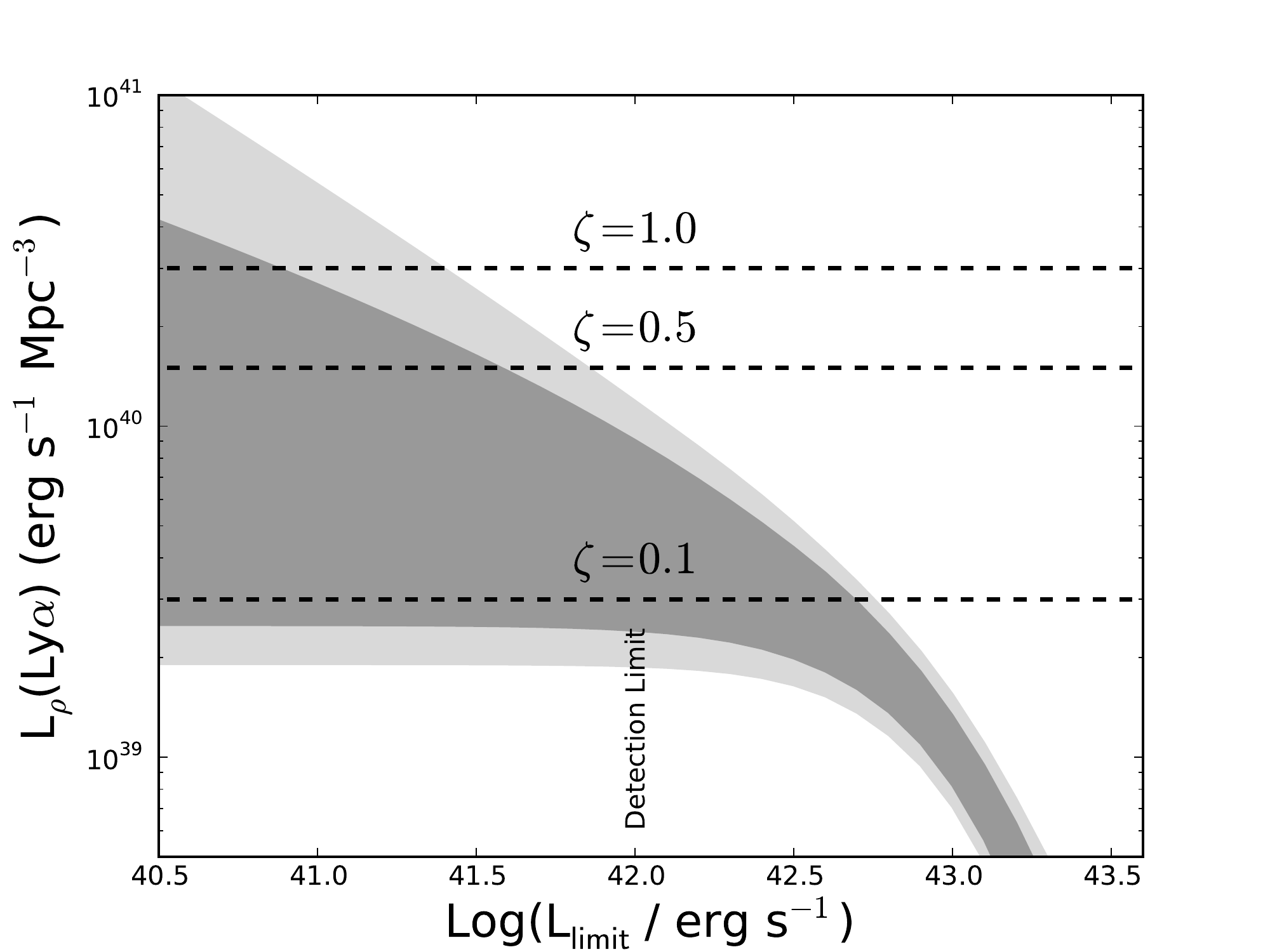}
\caption{Including galaxies fainter than those that we have observed increases the \lya\ luminosity 
density, making it easier for galaxies to maintain an ionized IGM.  Shaded regions show the 68 and 95\% confidence limits on the \lya\ luminosity
 density, as a function of the low-luminosity cutoff.   To simplify, the upper bounds are taken from 6 LAE 
 constraints and the lower bounds are taken from the 3 LAE constraints.    The critical \lya\ luminosity density 
 required to maintain an ionized IGM at redshift 5.7 is  
 $L_{\rho}(Ly\alpha) = \zeta \times 3.0\times 10^{40}$ erg s$^{-1}$ Mpc$^{-3}$.  
 Horizontal lines mark $\zeta = 0.1, 0.5,$and 1.0.   } 
\label{reion_fig} 
\end{figure} 

At $z\sim6$ the reionization of the intergalactic medium was complete, and the neutral
 hydrogen fraction in the IGM was negligible.  Star-forming 
galaxies are usually assumed to produce the necessary ionizing background to prevent 
hydrogen recombination.  However, observational 
constraints at redshifts 6 and higher have yet to demonstrate that galaxies
 produce the requisite ionizing background  
\citep{Bunker04, Bunker10, Bouwens07, Bouwens10, Oesch}.     Often,
 galaxies which are too faint to be observed are invoked \citep{YH04, Salvaterra, Finlator11}, 
 extrapolating faint counts from brighter populations.     In this paper, we address this shortcoming
  by detecting some objects from this hitherto undetected, postulated faint population (as demonstrated in \S \ref{lfsim_sec}).  

In Figure \ref{reion_fig}  we show the cumulative \lya\ luminosity density 
(as a function of the low-luminosity cutoff) allowed by our newly derived
 constraints.  
In order to determine whether LAEs can maintain an ionized IGM, we 
must infer an ionizing photon rate from the observed \lya\ photons, and
 then address whether this ionization rate can balance the recombination
  rate in the IGM.   This threshold depends on the escape of both \lya\ and hydrogen-ionizing (LyC)
  photons ($f_{LyC}$ and $f_{Ly\alpha}$; \citealt{Siana10, Nestor, Hayes11}),  as well as  the clumpiness of ionized
   hydrogen in the IGM ($C \equiv \langle n^2_{\textrm{\ion{H}{2}}}  \rangle / \langle n_{\textrm{\ion{H}{2}}} \rangle^2$).  In \clm\ and \dressler\ we renormalized these uncertain parameters to their best guess values and 
   grouped them into a single parameter, $\zeta$.  In this way, $\zeta = 0.9$ when $C=6$, $f_{Lya} = 0.5$, and $f_{LyC} = 0.1$, and the critical \lya\ luminosity density required to maintain
    an ionized IGM at redshift 5.7 is  $L_{\rho}(Ly\alpha) = \zeta \times 3.0\times 10^{40}$ erg s$^{-1}$ Mpc$^{-3}$.  
    Thresholds with $\zeta$ = 0.1, 0.5, and 1.0 are shown  in Figure \ref{reion_fig}.  This  plot
     illustrates that, if $\zeta$ is among the smaller
      values of 0.1 to 0.5, it may be possible to 
maintain an ionized IGM with only the galaxies that we have
 observed.   However, if $\zeta$ is larger, even fainter galaxies 
 are needed {\it and} the faint-end slope must fall among the
  steeper values allowed by our data.    As we pointed out in \dressler, when 
  faint-end slopes are shallow ($\alpha \gtrsim -1.3$), the \lya\ luminosity density
   grows very slowly when increasingly lower luminosities are included.   Therefore, in the case of a 
   shallow slope, it may be difficult for star-forming galaxies to reionize the IGM.

\subsection{Properties of galactic building blocks at the end of reionization} 
In \dressler\ we compared our density of faint LAEs to that of local galaxies,  estimating that we
 had found three or four sites of active 
star-formation per $L^*$ galaxy today.    The implication of this result was that we had identified the 
building blocks of Milky-Way 
type galaxies.  In this work, we have refined our faint-end slope measurements, and shown that the 
most-likely slope is shallower 
than the one we adopted in \dressler.   As a result, the space density of LAEs that we have identified
 is about a factor of two smaller--
$2\times 10^{-3}$ Mpc$^{-3}$.    However, not all galaxies have \lya\ emission.  By comparison, the
 space density of observed $i-$dropouts is about 2-3 times higher than the LAE density that we observe. 
  Furthermore,  if $i-$dropouts were detected as faint as $M_{UV} = -16.5$ (our estimate of our faintest continuum
   luminosities from \S \ref{lfsim_sec}), their space density would be several times higher than the LAE density
    presented here.   Since our LAEs belong to this class of postulated ultra-faint $i-$dropouts,  we can infer that we
     have observed the building blocks of present day $L^*$ galaxies.

While it remains difficult to determine the physical properties of LAEs from their \lya\ emission line 
alone, in \dressler\ we placed 
their properties in the larger context by comparison to other studies.  For example, the angular 
correlation of the brighter LAEs presented in \cite{Ouchi08,Ouchi10} implies  halo 
masses of $M = 10^{11}- 10^{12}~ M_{\sun}$ at redshift 5.7.    Presumably, the  lower
 luminosity objects that we have found reside in 
less massive halos.   Additionally, in \clm\ we estimated SFRs as low as 1 to a few 
$M_{\sun}$ yr$^{-1}$ (if half 
of the \lya\ photons escape the galaxy).  If this star formation rate persists 
for a dynamical time scale of at least 50 Myr, then approximately
 $10^{8}$ $M_{\sun}$ of stars will have formed.   These masses, SFRs, and redshifts
  are well matched to the predicted properties of the galaxies that enriched the IGM with metals  \citep{Opp09, clm10}. 

As noted in \S \ref{confirmed_sec}, some of the LAEs have spatially resolved emission in our DEIMOS
spectroscopy.  This is not strictly surprising, as stacking of narrowband \lya\ images
at $z\sim 2-3$ has revealed that diffuse, extended \lya\ emission is a generic component LAEs and 
LBGs alike \citep{Steidel11}.  Additionally,  \cite{Rauch08} find that, when  exceptionally faint
 surface brightness limits are reached, a 
considerable fraction (40-70\%) of redshift 3 LAEs are extended.  Consequently, the observation that
 LAEs are compact in narrowband HST images  \citep{Bond10, Finkelstein11}   does not
  imply that the \lya\ emission region is small, but rather that the sensitivity to low surface 
  brightness emission is insufficient in the 
space-based images.   Further studies of extended \lya\ emission in larger samples may help 
quantify  the evolution of the circumgalactic medium at high-redshift.

\section{Summary and Conclusions} 
We have measured the faint-end slope of the \lya\ luminosity function 
at redshift 5.7, using Multi-slit Narrowband 
Spectroscopy (MNS) to identify faint emission lines.  In \dressler\ \citep{Dressler11}, 
we presented over 200 \lya\ candidates and showed that a steep faint-end slope ($\alpha\sim-2$) was
 likely.  Now, we have added spectroscopic followup to distinguish foreground
line emitters and confirm LAEs.  In this way, we have conclusively identified three, and tentatively identified 
three 
further LAEs.   With these LAEs in hand, we develop the appropriate framework for deriving a luminosity
 function 
from blind-spectroscopic data, where the positions of galaxies within the search slit are not always known.   
The resulting 
\lya\ LF has a faint-end slope of $-1.70^{+0.73}_{-0.57}$  ($-1.45^{+0.92}_{-0.70}$) when six (three) LAEs are 
used.   When our spectroscopic followup in COSMOS and the LCIRS 15H field is complete, 
we can expect to identify around 40 LAEs, and improve the uncertainties on the LF measurements.

Studies of galaxies in the post-reionization era ($z\sim4-6$) allow us to map 
the growth of young objects
in an early universe.    For the LAEs in our survey, we estimate SFRs of one to a
 few $M_{\sun}$ yr$^{-1}$,  and a stellar mass of at least $10^{8}$ M$_{\sun}$. 
   In fact, these 
properties are matched to the masses, SFRs, and redshifts of galaxies that are thought to
 have enriched the IGM with metals \citep{Opp09, clm10}.

By identifying some of the faintest galaxies ever observed at $z\sim6$, we uncover objects 
that are important for the reionization of the IGM.    In \S \ref{reion_sec}, we 
show that by extrapolating our faint-end slope measurement to low luminosities, it may be
 possible  for LAEs to maintain an ionized 
IGM with our nominal assumptions about the escape fractions of LyC and \lya\ photons 
($f_{LyC}=0.1$ and $f_{Ly\alpha}=0.5$) and the clumpiness of the ionized hydrogen in the IGM ($C=6$).  
 Nevertheless, the allowed \lya\ luminosity density of sources brighter than $10^{41}$ erg s$^{-1}$ 
 spans more than an order of magnitude. If the true value falls among the lower range allowed by our data, 
 the escape fractions and clumping 
factor may need to be revised in order for galaxies to maintain an ionized IGM.   Ultimately, this requirement 
may prove reasonable-- the escape fractions and clumping factor are all uncertain.   Future efforts to constrain 
these parameters, in parallel with improved measurements of galaxies will 
be needed if we are to conclusively determine, whether star-formation at $z\sim6$ maintains the ionized IGM.

The LAEs presented here set a benchmark for comparison to 
higher redshift \lya\ surveys that aim to measure the reionization of the
 IGM.   By combining our line width measurements with those reported by
  \cite{Hu10}, we 
confirm that fainter LAEs have somewhat narrower line widths. 
 Although the origin of this correlation is unclear, it must
be taken into consideration when drawing conclusions about a
 (partially) neutral IGM from \lya\ line profiles, since higher-redshift surveys will 
naturally be limited to more luminous LAEs.   Furthermore, our 
measurement of the faint-end slope of the \lya\ LF represents the first
step towards measuring the faint-end evolution imparted by a neutral IGM.  
 An MNS survey 
at redshift 6.5 could determine if the evolution observed at the bright-end 
 \citep{Kashikawa06, Kashikawa11, Ouchi10} persists at the faint-end.  
 Such observations may constrain models that describe the propagation 
 of ionized regions in a partially neutral IGM (e.g.\ \citealt{ME2000, CM03,Furlanetto2006, Ono}).  

In summary, we have shown that spectroscopic searches for LAEs are able to uncover
 faint galaxies that cannot be observed
by any other means.    Measuring this population opens new windows to understanding
 the formation of young galaxies, and the 
enrichment and reionization of the IGM.

\acknowledgements
The authors wish to acknowledge the Subaru Deep Field team, 
namely Nobunari Kashikawa, Matt Malkan,  and Kazuhiro Shimasaku
for providing data necessary to complete this paper. Similarly, the high-level 
science products provided by the COSMOS team have been invaluable. 
In addition, critical support at various stages of this project have come  
from Matt Auger, Peter Capak, Michael Cooper, Keith Farley, Kristian Finlator, Moire Prescott,  
Anna Nierenberg, and an anonymous referee. 
This research has made use of the NASA/ IPAC Infrared Science Archive, 
which is operated by the Jet Propulsion Laboratory, California Institute of Technology, under 
contract with the National Aeronautics and Space Administration.   C.\ L.\ M. thanks the David and Lucile 
Packard foundation for support. 
The authors wish to recognize and acknowledge the very significant cultural role and reverence that
 the summit of Mauna Kea has always had within the indigenous Hawaiian community.  We are most fortunate 
 to have the opportunity to conduct observations from this mountain.

\appendix 
\section{A. Survey Completeness} 
\label{comp_app}
\begin{figure} 
\plottwo{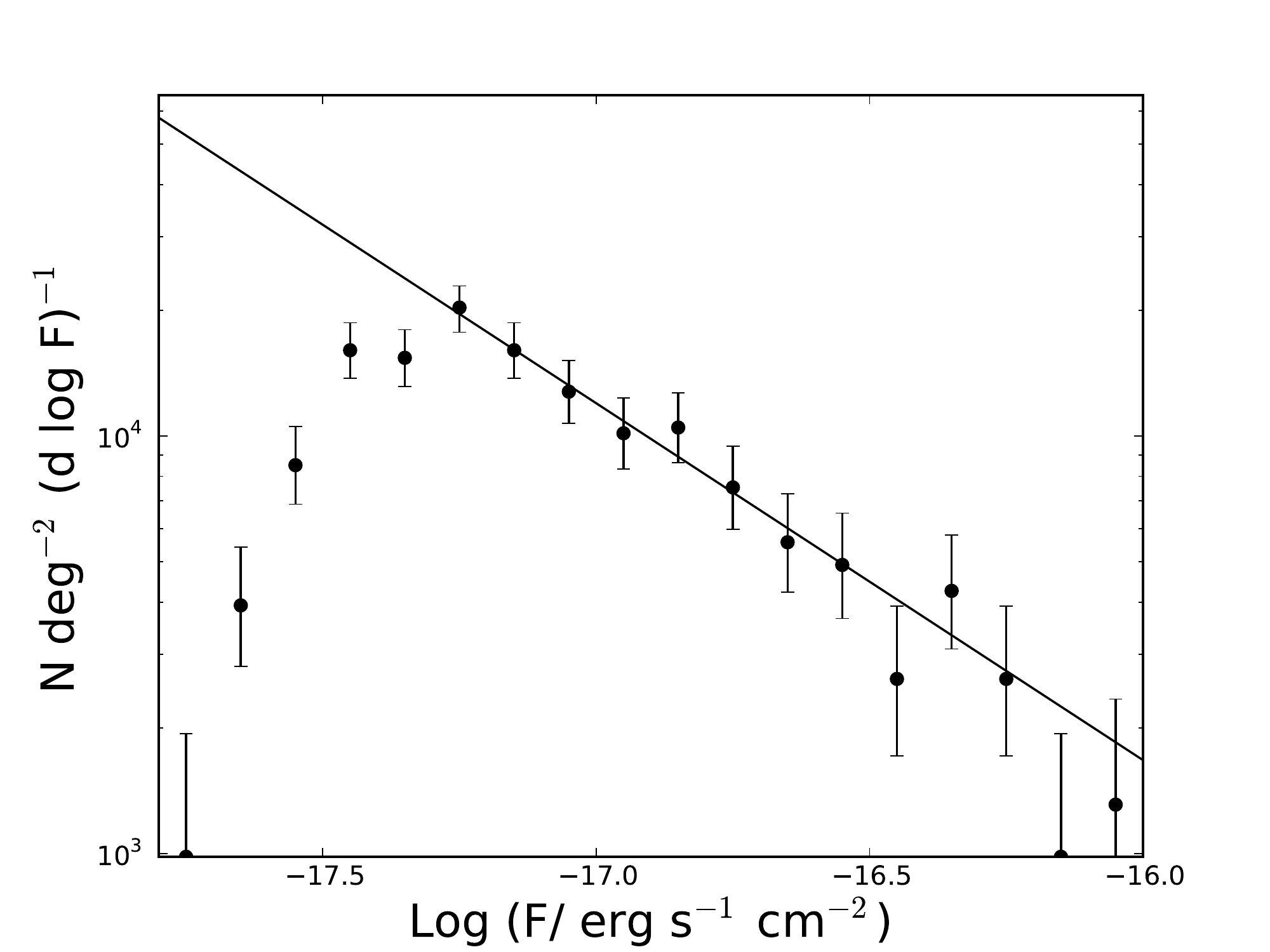}{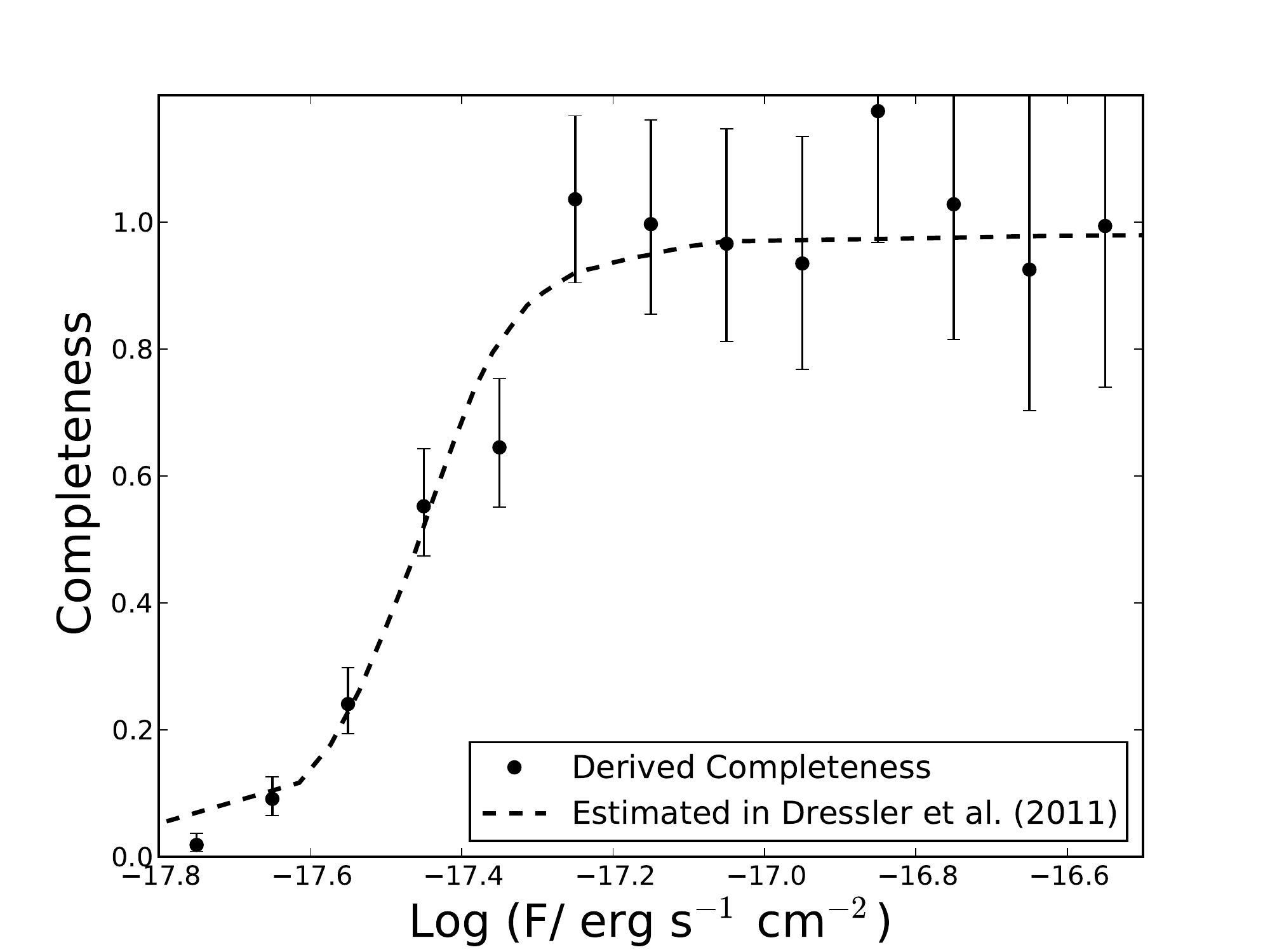} 
\caption{The source counts of all emission line galaxies in our survey follow a power law, and are highly complete brighter than $F= 10^{-17.2}$ erg s$^{-1}$ cm$^{-2}$.  A fit to the counts above this threshold gives a power law index of $-1.02 \pm 0.12$.  Comparison of the observed counts to the extrapolated fit gives the survey completeness, shown on the right.    } 
\label{comp_fig} 
\end{figure}

To calculate the survey incompleteness for the IMACS search data, we take advantage of the fact that our emission line counts are 
well described by a power law between $F = 10^{-16}$ and $10^{-17.2}$ erg s$^{-1}$ cm$^{-2}$ (see Figure \ref{comp_fig}) .     This is easily understood, because we are  summing the faint-end counts of  (primarily) the constituent foreground populations-- all of which are described by power-laws.   Clearly, we
can expect these counts to continue to rise to fainter emission line fluxes.  Therefore, we predict the the number of all emission line galaxies by fitting  power-law to the counts brighter than $10^{-17.2}$ erg s$^{-1}$ cm$^{-2}$  in both the COSMOS and 15H fields.  
(The sensitivity is the same for both fields, so we include the 
15H field here to improve statistics.)     This fit, illustrated in Figure \ref{comp_fig} goes as $F^{-1.02 \pm 0.12}$.  The completeness of our survey-- defined as the ratio of the observed counts to those predicted by our model-- is plotted in Figure \ref{comp_fig}.   A completeness fraction slightly greater than unity naturally arises when the power-law fit falls slightly below the data. 

For comparison, we show the model adopted by \dressler, which was estimated from a modification to the completeness function presented in \clm.   The good agreement
between the estimate in \dressler\ and the completeness calculated here  indicates that no substantial revisions to the implications in \dressler\ are needed.  
  Additionally, we note that the faintest bin that we consider in this paper contains many galaxies (42), so the Poisson noise on this calculation makes a negligible contribution 
to our error budget.   Finally, while one might expect that the completeness to line emission is different for LAE candidates (without detected continuum)  and foreground galaxies (with detected continuum),  the faintest bins are dominated by LAE candidates, so it is reasonable to apply this incompleteness correction to the  LAEs in our survey.

 \section{B. Foreground Populations} 
 
\label{fgsec} 
In \dressler, we characterized the foreground \ha, \oiii, and \oii\ populations that comprise our sample of emission lines.  Subtracting 
the foreground LFs estimated from the COSMOS field showed a significant excess of counts that suggested a steep rising slope ($\alpha \sim -2$) for the remaining galaxies that are likely LAEs.   Additionally, an angular correlation of the \lya\ candidates and \oii\ emitters  showed that fewer than 30\% of the LAE candidates should be \oii.  Since even fewer should be \ha\ and \oiii, the implication is that approximately 100 LAEs may exist in the combined COSMOS and 15h survey fields.   
In this section, we show that the foreground number counts  in COSMOS are indeed higher than what we adopted in \dressler, consistent with our preference for a slightly shallower faint-end slope for the \lya\ LF. 

  In \S \ref{ratesec} we identified foreground galaxies through two means:  objects which are obvious foreground emitters in the search data, and objects which were 
  initially LAE candidates in the search data, but are now spectroscopically or photometrically identified as foreground galaxies.    
  The counts of the total foreground populations are  listed in Table \ref{counts1}, and shown again, for reference in Table \ref{foregrounds}  where we compare these counts to the numbers predicted by the foreground luminosity  functions used by \dressler.   In each bin, the foreground galaxy counts that we have identified {\it so far} are similar to (often exceeding) the number predicted by foreground LFs.   In addition, in \S \ref{ratesec} we showed that the results of our spectroscopic followup indicate that the 
unidentified LAE candidates (for which counts are also listed in Tables \ref{counts1} and \ref{foregrounds}) are predominantly foreground galaxies.  
Consequently, we expect that when our spectroscopic followup is complete, the observed foregrounds will be higher than the model favored in \dressler.
We estimate these counts in the same way that we calculated the inferred LAE counts in \S \ref{ratesec}.  The inferred foreground number counts are given for two cases, with the low (high) counts referring to the case of six (three) genuine LAEs in this paper. 

Including the foreground galaxies  inferred among the unidentified line emitters makes the case clear: the nominal ``best estimate''  foreground counts  in \dressler\ were underestimated, at least for the COSMOS field.  However, they remain within range of reasonable foreground LFs which we considered.  An alternative model is shown in Table \ref{foregrounds} where the faint-end slopes are all steeper by $1\sigma$ from their adopted values (see \dressler).  These counts agree more closely with our expectation from followup spectroscopy.     However, a model where all three foreground populations have steeper slopes by 1$\sigma$ is contrived.  It is more probable that the number counts  of one of the foreground populations has been  significantly  underestimated.

\begin{deluxetable*} {ccccccc}
\tablecolumns{7}
\tablecaption{Foreground Number Counts}   
\tablehead{
\colhead{Flux} & \colhead{Identified Foregrounds} & \colhead{Unidentified }  & \colhead{ $N_{pred}$\tablenotemark{a}} & \colhead{Inferred (Low)} &  \colhead{Inferred (High)} & \colhead{$N_{pred}$ (+ 1$\sigma$)\tablenotemark{a} }  
}
\startdata
-17.47 &   19 & 22  & 23.0 &   38.9  & 42.0 & 33.6  \\
-17.27 &   39 &  13 & 35.8 & 47.8 & 49.9  & 49.0   \\
-17.07 &   33  &  9  & 30.0 & 38.2 &  40.8 & 38.3   \\
-16.87 &   24  &  6  & 22.5 & $\sim30$ & $\sim 30$ & 27.0 \\
-16.67  &  14  & 2   & 16.5 &$\sim16$  & $\sim 16$ &  18.5 \\
-16.47  &   14  &  0   & 10.8  & 14 & 14  & 11.3   \\
-16.27 &  13  &  0   & 6.2 &  13 & 13  & 6.3 \\
-16.07 &    6  &  0  &  3.0 & 6 & 6  & 3.0 \\
-15.87 &   2  &  0   & 1.4 & 2 & 2 & 1.4  \\
-15.67 &   1  &  0   &  0.7 &  1 & 1 & 0.7
 \enddata
\label{foregrounds} 
\tablenotetext{a}{Model predictions of the observed foreground galaxy counts in include the incompleteness given in Appendix \ref{comp_app}.  }
\end{deluxetable*}

The breakdown of the spectroscopically identified foreground galaxies in Table \ref{rates} suggests  that \hb\ emitters are the cause of the larger foreground counts.   In \cite{Dressler11}, this 
foreground population was not explicitly considered.   It was, however, implicitly included with the \oiii\ population.   When foreground 
emission line galaxies are detected through narrowband imaging, broad-band colors are typically used to identify the approximate redshift and probable line \citep{Takahashi, Ly, shioya}.  Not only does this mean that the \oiii\ catalog will contain some \hb\ emitters; an \ha\ catalog may be contaminated by 
[\ion{S}{2}] 6716, 6730 \AA\ emitting galaxies.   In practice, spectroscopic followup of bright emission lines from narrowband imaging samples has shown that contamination of this form is rare (Ly et al.\ 2007 
find 16 \oiii\ emitters for only two \hb\ emitters).   Even if this contamination were significant, the overall \oiii\ + \hb\ luminosity function should be sufficient to allow for an accounting of the total foreground populations {\it at the bright end}, as we did in \dressler.  

\begin{figure} 
\plottwo{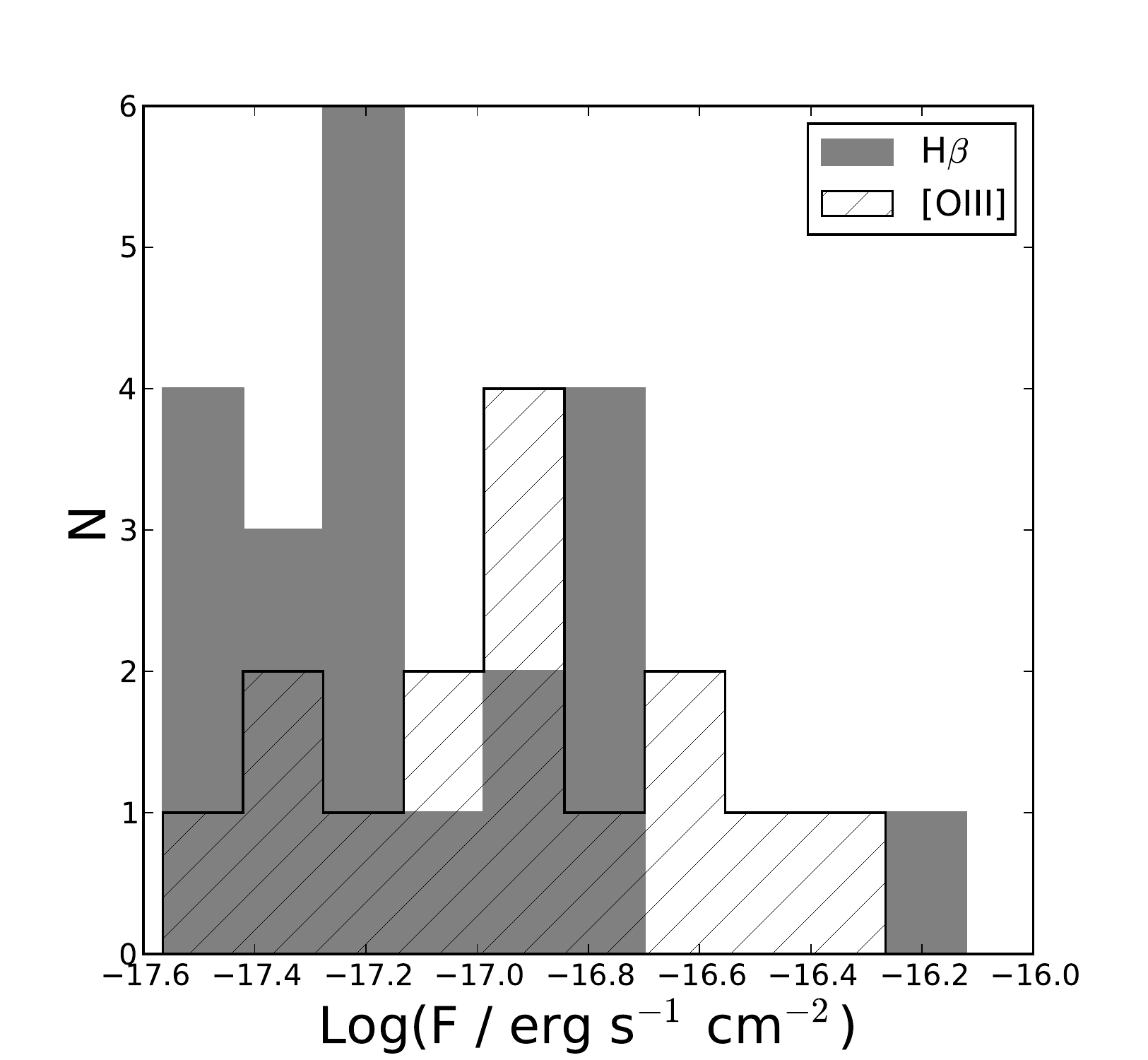} {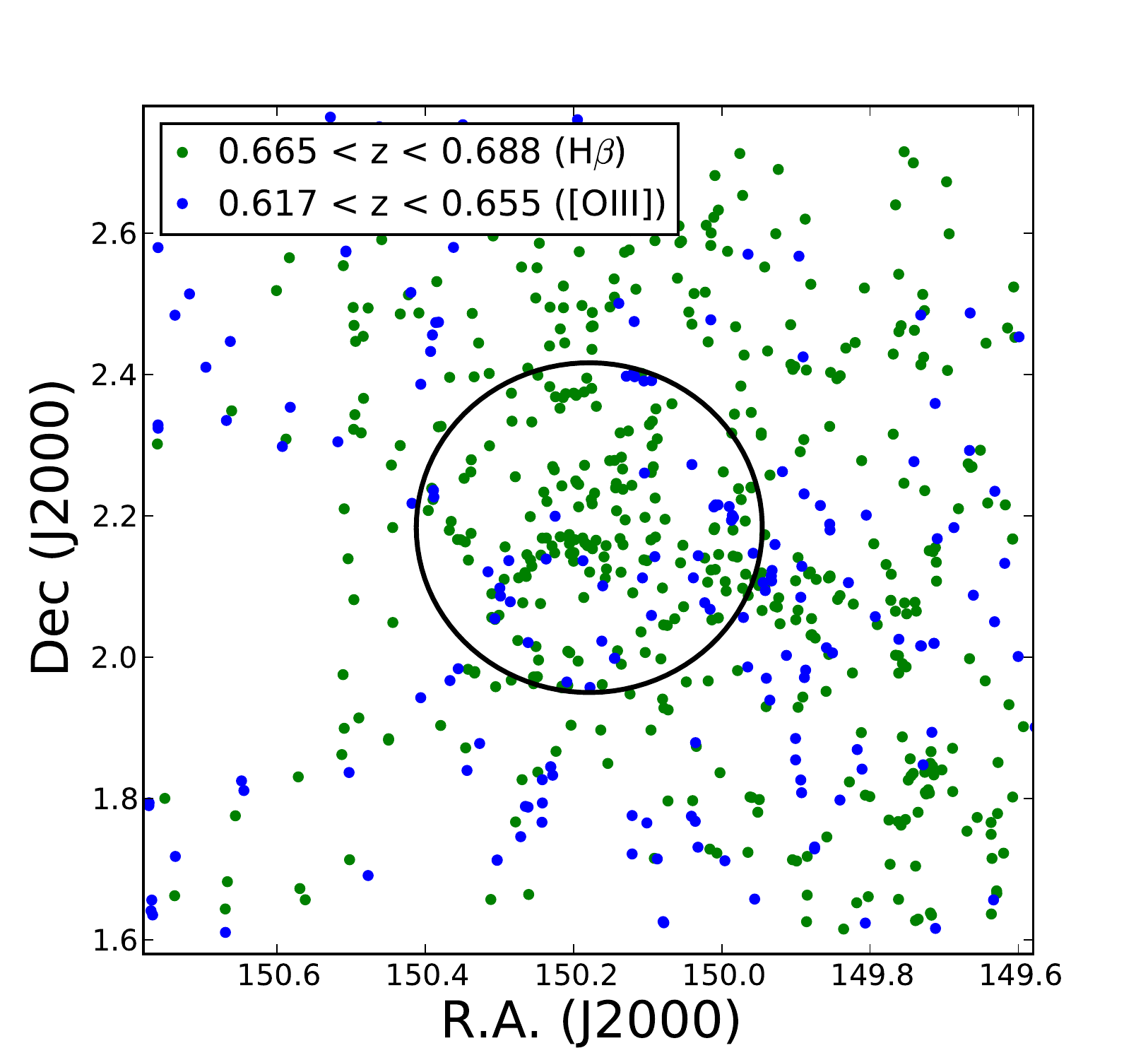}
\caption{{\it Left--} Spectroscopically confirmed \hb\ emitters outnumber \oiii\ emitters at faint fluxes. {\it Right-}   The zCOSMOS (bright) spectroscopic catalog \citep{Lilly}
contains a significantly higher density of galaxies at $z\sim0.68$ than at $z\sim0.63$.  Objects are shown for the redshift windows that correspond to 
\hb\ and \oiii $\lambda \lambda$4959, 5007 in our survey, although the latter volume is 50\% larger.   The footprint  of our IMACS field is shown 
by the black circle.  Within this region, there are four times as many galaxies at $z\sim0.68$ than at $z\sim0.63$, implying a volume density six times higher.  We note that the coverage of zCOSMOS is not uniform over the regions shown here, but the relative numbers of the green and blue points should be 
unaffected.  } 
\label{hbfig}
\end{figure}

Our spectroscopic followup observations, which included known foreground objects as slit-mask filler, provide 
strong evidence that \hb\ emitters in COSMOS play a more predominant role than we assumed in \dressler. 
In total, we have identified 15  \oiii\  and 21 \hb\ emitters.    The distributions of their  fluxes are shown in Figure \ref{hbfig}, where, relative to the numbers of \oiii\ emitters,  an excess of   \hb\ emitters fainter than about $8 \times 10^{-18}$ \flux\ is observed. 

One way to interpret the excess of faint \hb\ (relative to \oiii) emitters is as an indication that
the faint-end of the \hb\ LF is significantly steeper than the \oiii\ LF.  This would be the case if
fainter \oiii\ emitters have progressively lower \oiii/\hb\ ratios.    Since \oiii/\hb\ is strongly correlated 
with the R23  (\{\oiii  + \oii\}/\hb; \citealt{Pagel})  metallicity diagnostic, low \oiii/\hb\ ratios could be an indication of
low oxygen abundance (when galaxies are on the lower-branch of R23).  In practice, however,  even 
the lowest metallicity galaxies observed at $z<1$ have high \oiii/\hb\ ratios (larger than three;  \citealt{Salzer, Kakazu}).  The \oiii/\hb\ ratios that we measure in our DEIMOS observations are similarly high.    This means
that it is difficult to interpret the excess \hb\ counts as an indication of low-metallicity galaxies.

A more probable explanation for the \hb\ emitters is that we have observed an overdensity at $z\sim0.68$. 
In Figure \ref{hbfig}, we use the zCOSMOS (bright) catalog of spectroscopic redshifts \citep{Lilly} to show the spatial distributions of galaxies in the redshift
windows that corresponds to   \hb\ and \oiii $\lambda \lambda$4959, 5007.  Even though the volume at the ``\oiii\ redshift'' is 50\% larger than  
at the ``\hb\ redshift'', the full zCOSMOS catalog contains 2.5 times as many galaxies in the latter redshift window.  
 Within the footprint of our IMACS field (see Figure \ref{hbfig}, right),  this excess density is amplified, with a factor of four more galaxies in the redshift window at $z\sim0.68$ than at $z\sim0.63$.   The excess of  faint \hb\ emitters in Figure \ref{hbfig} (left) is easily explained by this overdensity, 
 which is most apparent at luminosities fainter than the knee of the \hb\ LF.  Therefore, the COSMOS foreground counts at faint fluxes are indeed 
 higher than the model adopted in  \dressler.

 This increase over our previous estimates of the foreground populations in COSMOS  does not erase our signal from LAEs.   Although we find a slightly shallower slope ($-1.7$ as opposed to $-2.0$), these results are consistent, as $\alpha = -2.0$ falls within the 68\% confidence interval that we present in this paper.    Further followup spectroscopy (including our 15H field) will constrain the faint-end of the \lya\ LF more tightly and simultaneously measure the faint foreground populations.

\end{document}